%% file: ms.tex
\documentclass[a4paper,11pt]{amsart}
\input{preamble}

\begin{document}


\title{Multiscale methods for signal selection in single-cell data}

\author[R. S. Hoekzema, L. Marsh, O. Sumray, T. M. Carroll, X. Lu, H. M. Byrne, H. A. Harrington]{Renee S. Hoekzema$^{*,1,2}$, Lewis Marsh$^{*,1,3}$, Otto Sumray$^{*,1,3}$, Thomas M. Carroll$^{3}$, Xin Lu$^{3}$, Helen M. Byrne$^{1,3}$, Heather A. Harrington$^{1,4}$}
\address{$^1$Mathematical Institute, University of Oxford\\
$^2$Department of Mathematics, Free University of Amsterdam\\
$^3$Ludwig Institute for Cancer Research, University of Oxford\\
$^4$Wellcome Centre for Human Genetics, University of Oxford
}

\email{r.s.hoekzema@vu.nl}

\date{\today}

\begin{abstract}
Analysis of single-cell transcriptomics often relies on clustering cells and then performing differential gene expression (DGE) to identify genes that vary between these clusters. These discrete analyses successfully determine cell types and markers; however, continuous variation within and between cell types may not be detected.
We propose three topologically motivated mathematical methods for unsupervised feature selection that consider discrete and continuous transcriptional patterns on an equal footing across multiple scales simultaneously.
Eigenscores ($\eig_i$) rank signals or genes based on their correspondence to low-frequency intrinsic patterning in the data using
the spectral decomposition of the Laplacian graph. 
The multiscale Laplacian score (MLS) is an unsupervised method for locating relevant scales in data and selecting the genes that are coherently expressed at these respective scales.
The persistent Rayleigh quotient (PRQ) takes data equipped with a filtration, allowing the separation of genes with different roles in a bifurcation process (e.g., pseudo-time).
We demonstrate the utility of these techniques by applying them to published single-cell transcriptomics data sets. The methods validate previously identified genes and detect additional {biologically meaningful} genes with coherent expression patterns.
By studying the interaction between gene signals and the geometry of the underlying space, the three methods give multidimensional rankings
of the genes and visualisation of relationships between them.
\end{abstract}

\maketitle
\blfootnote{$^*$These authors contributed equally}

\input{introduction}

\input{materials_and_methods}

\input{results}

\clearpage

\input{conclusions}

\section{Acknowledgements}
The authors thank Mariano Beguerisse, Carla Groenewegen, Joe Kaplinsky, and Vidit Nanda for helpful discussions. We thank Renaud Lambiotte and Michael Schaub for reading an earlier version of this manuscript.
HAH gratefully acknowledges funding from EPSRC EP/R018472/1, EP/R005125/1 and EP/T001968/1, the Royal Society RGF$\backslash$EA$\backslash$201074 and UF150238. RSH, HAH and HMB acknowledge funding from the Emerson Collective. This research was funded in part by EPSRC EP/R018472/1.
LM, OS, TMC, XL and HMB are funded by the Ludwig Institute for Cancer Research Ltd.
TMC gratefully acknowledges scholarship support from the Rhodes Trust.
For the  purpose of Open Access, the authors have applied a CC BY public copyright licence to any  Author Accepted Manuscript (AAM) version arising from this submission.

\clearpage
 
\input{additional_figures}

\clearpage

\printbibliography

\end{document}

%% file: preamble.tex
\usepackage[margin=1.2in]{geometry}
\usepackage{adjustbox}

\graphicspath{ {./images/} }
\usepackage{amsmath,amssymb,amsfonts,amsthm}
\usepackage{eqnarray}
\usepackage{tikz, tikz-cd}
\usepackage{caption, subcaption} 
\usepackage{enumerate}
\usepackage{geometry}
\usepackage{verbatim}
\usepackage{graphicx} 

\usepackage{thmtools}
\usepackage{thm-restate}
\usepackage{todonotes}
\usepackage{inputenc}
\newcounter{todocounter}

\usepackage{appendix}

\newcommand{\RR}{\mathbb{R}}
\newcommand{\LL}{\mathcal{L}}

\DeclareMathOperator{\eig}{eig}

\newcommand{\ac}{\alpha^c}
\DeclareMathOperator{\PRQ}{PRQ}
\newcommand{\NPRQ}{\widehat{\PRQ}}

\newtheorem{lemma}{Lemma}

\theoremstyle{definition}
\newtheorem{definition}[lemma]{Definition}
\theoremstyle{remark}
\newtheorem{remark}{Remark}
\newtheorem*{remark*}{Remark}

\usepackage[backend=biber, date=long, giveninits=true, terseinits=true, maxnames=10]{biblatex}
\bibliography{ms.bib}

\usepackage{nameref}
\usepackage{enumitem}
\DeclareNameAlias{sortname}{family-given}
\DeclareNameAlias{author}{family-given}

\usepackage[hidelinks]{hyperref}

\newcommand\blfootnote[1]{%
	\begingroup
	\renewcommand\thefootnote{}\footnote{#1}%
	\addtocounter{footnote}{-1}%
	\endgroup
}

\usepackage{appendix}

%% file: introduction.tex
\section{Introduction}

Cells, the building blocks of life, are often classified into discrete cell types (e.g., liver, neuron, immune, or blood cells). 
In modern experiments, cell type identification commonly relies on partitioning single cell RNA sequencing (scRNA-seq) data.
Differential gene expression (DGE) algorithms use statistical tests to determine genes that significantly differ between predefined groups of cells.  However, cellular biology is more nuanced: there are multiple scales of cell classification (e.g., Treg cells are T cells which are a type of immune cell), continuous transitions into cell types (e.g., embryonic development starts from stem cells that differentiate into broad cell types that further specialise), or natural variations within cell types. The rich repertoire of gene expression patterns and cellular subphenotypes offers an opportunity to study continuity of gene expression. \par

Mathematically, single-cell data are given as raw counts of RNA transcripts that represent the 
expression of more than 20,000 genes in the human genome. A cell-by-gene matrix of counts is then preprocessed to reduce noise, variance due to technical effects, and the number of genes to form a smaller normalised gene expression matrix $\widehat{Y} \in \mathbb{R}^{m \times n}$,
where $m \sim 10^{3}$ genes and $n \sim 10^{3}$--$10^{6}$ cells~\cite{SeuratV4,wolf2018scanpy}. 
Due to the high-dimensional nature of these data, along with sparsity and noise, standard data science methods are out of reach. 
\par
The field of topological data analysis (TDA) studies the shape and connectivity of data at multiple scales of resolution. TDA methods require a metric and approximate the shape of the data by building covers or sequences of higher order networks (i.e., filtrations) on the data. 
TDA methods have successfully analysed and visualised single-cell data (e.g., UMAP, which relies on fuzzy simplicial sets; or Mapper, which visualises data using covers and filters)~\cite{mcinnes2018umap, becht2019dimensionality,jeitziner2017two,rizviSinglecellTopologicalRNAseq2017,kuchrooTopologicalAnalysisSinglecell2021,vandaeleStableTopologicalSignatures2021}. 
In this paper, instead of studying the shape of data, we focus on the related task of quantifying how well signals on a given data set align with the topology of the data.

\par

The multiscale nature of topological data analysis and filtrations leads us to combine these ideas with graph signal processing~\cite{ortega2018graph} and spectral graph theory~\cite{chung1997spectral} to study the continuous variation of gene features across cells.
The analysis in this paper starts with a preprocessed single-cell data matrix $\widehat{Y}$, as computed in the standard software Seurat~\cite{SeuratV4}, and then, it uses UMAP to construct an undirected weighted $k$-nearest neighbour cell similarity graph $G$. The nodes, which represent cells, are connected by edges and weighted by the similarity of gene expression. 
{ On this graph, the expression of a single gene is now a 
real-valued function $g: V \to \RR$ on the vertices of the graph, which is known as a graph signal in
graph signal processing.
Viewing a gene as a function on vertices of a graph, or as a graph signal, is implicit when applying DGE to a clustering of graph vertices.}

\par 
Spectral graph theory, graph signal processing and the emerging field of topological signal processing~\cite{robinson2014topological,schaub2021signal, barbarossa2020topological} offer a setting to study
and {compare continuous patterns of functions across nodes, edges, and higher-order data structures.
He et al. introduced the Laplacian score in~\cite{He2005} as a method for feature selection
on point cloud data. Feature selection in machine learning is a process of dimensionality reduction
of data before other analyses are applied.
Govek et al.~\cite{govekClusteringindependentAnalysisGenomic2019}
applied and extended the Laplacian score to single-cell data
as an improvement on using the variance of gene expressions
to rank genes by importance.}
By studying the spectral properties of the Laplacian graph, each 
gene is given a score according to its consistency with the local geometric structure of the graph~\cite{govekClusteringindependentAnalysisGenomic2019}. The Laplacian score is small if the gene signal roughly correlates with the graph structure (i.e., it is locally approximately constant but has global variation) and is large if the expression of a gene varies wildly on local neighbourhoods.  
This feature selection approach ranks the best features (e.g., genes) from the input data to form a compact and informative data representation. 
A score for each gene can be calculated, providing an overall ranking of features or gene signals ~\cite{He2005,govekClusteringindependentAnalysisGenomic2019}. 

\par

\begin{figure}[h]
 
	\includegraphics[width=0.8\textwidth]{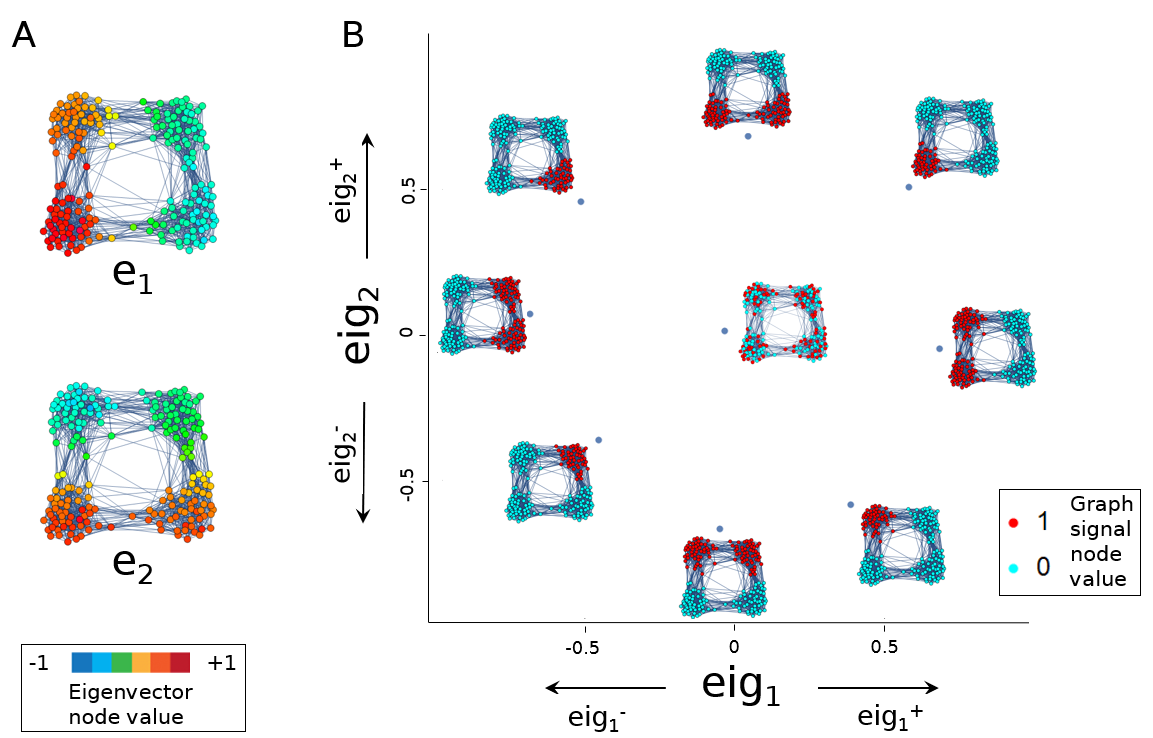}
	\caption{{{The} eigenscore method 
	(defined in Section \ref{section:MM_eig}) demonstrated here on a graph constructed by taking 100 random points each from of four touching balls in 30 dimensions and connecting them via a 15-nearest-neighbour graph. (\textbf{A}) Laplacian eigenvectors $e_1$ and $e_2$ distinguish the left and right two clusters and the top and bottom two clusters, respectively. (\textbf{B}) Different graph signals align or anti-align differently with the two eigenvectors, resulting in a plot of eigenscore $(\eig_1$, $\eig_2)$-space that differentiates the various signals. A random signal plots near the origin.}}
	\label{fig:eig_explain}
\end{figure}

In this work, we analyse gene signals on a cell similarity graph, while taking into account multiple scales of the single-cell data. We propose three computationally tractable methods for finding gene expression patterns that drive continuous variation in the data set. 
Similar to Govek et al.~\cite{govekClusteringindependentAnalysisGenomic2019}, the proposed methods do not require clustering cells or predefined cell assignments.  {Therefore, the selected gene signals  
are agnostic to cell types or  clusters, 
gene selection 
is 
continuous across the cell network and some genes may link across multiple `cell types'. In this sense, DGE based on standard clustering approaches will only identify genes that are specifically enriched in a pre-specified cell population, whereas the proposed methods can identify genes common to disparate cell types.} Instead of one ranking, we propose multiple rankings of gene expression patterns at different scales of the data. Briefly, eigenscores
restrict the signal to the eigenspaces corresponding to the smallest eigenvalues. We score each gene by its alignment to each of the eigenvectors with the
smallest eigenvalues
and then visualise the signals in gene space (Figure \ref{fig:eig_explain}).
Our proposed multiscale Laplacian score (MLS) pipeline uses the theory of continuous-time random walks and Markov stability~\cite{Delvenne2013,Schaub2012} to rank genes according to their consistency with features that range from local to global geometric structures. 
The persistent Rayleigh quotient (PRQ) takes in a filtration on the data (e.g., time) to study bifurcation patterns in gene expression data. The PRQ is based on the Kron reduced (persistent) {Laplacian}~\cite{Dorfler2013,wangPersistentSpectralGraph2019a, memoliPersistentLaplaciansProperties2021}, which considers subgraphs inside a larger graph. It then applies the Rayleigh quotient associated with this operator, resulting in the identification of genes that drive bifurcation processes. To probe the discrete cell type paradigm, we apply the methods to synthetic and experimental data sets, which select subsets of genes that span known cell types and provide possible pathway transitions between them. 

The article is organised as follows.
In Section \ref{section:MM_SGT}, we present mathematical preliminaries. We then introduce the three proposed scores (Sections \ref{section:MM_eig}--\ref{section:MM_PRQ}) and data sets (Section~\ref{sec:data}). In Section \ref{section:Results}, we present and discuss the computational results, highlighting the potential of each method for application on single-cell data sets, and then, we conclude in the final section.

%% file: materials_and_methods.tex
\section{Materials and Methods}\label{section:MM}

\subsection{Preliminaries}\label{section:MM_SGT}

Let $ G = (V, E)$ be an undirected graph where $V=\{1,\ldots,n\}$ are nodes (representing the set of $n$ cells) and $E \subseteq V \times V$ are edges that are weighted by gene correlation or similarity. The weight between cells $u$ and $v$ is recorded in the $a_{uv}$ entry of the weighted adjacency matrix $A$.  Let $d_v$ denote the degree of node $v$, and let $D$ denote the diagonal matrix $D_{vv}$ whose entries have value $d_v$. 

\begin{definition}
The {combinatorial Laplacian} \(L\), the  {symmetrically normalised Laplacian} \(\LL\) and the {random walk Laplacian} $L^\mathrm{rw}$ of the graph $G$ are:
\begin{align}
L &= D - A,\label{eq:laplacian} \\
\LL &= D^{-1/2}L D^{-1/2} = I - D^{-1/2}AD^{-1/2},\label{eq:normalised_laplacian}\\
L^\mathrm{rw}&=D^{-1}L=I-D^{-1}A.\label{eq:rw_laplacian}
\end{align}
\end{definition}

\begin{definition}
The \textit{Rayleigh quotient} for a non-zero graph signal $g: V \to \RR$ on the nodes of \(G\) is
\begin{equation}
\label{eqn:rayleigh-quotient}
R_L(g)
= \frac{\langle g, L g \rangle}{\langle g, g \rangle}
= \frac{
    \sum_{u\sim v}{A_{uv}
    (g(u) - g(v))^2}
    }{
    \sum_{u}{g(u)^2}
    },
\end{equation}
where $u\sim v$ indicates that $u$ and $v$ are adjacent nodes in \(G\) and the inner product is defined as $\langle g  , h \rangle= \sum_{v\in V} g(v) h(v).$ 
\end{definition}
If \(g\) is constant, then \(R_L(g)\) is zero.
Substituting the normalised Laplacian into \mbox{Equation~\eqref{eqn:rayleigh-quotient}},
we have the following equation:
\begin{equation}
\label{eqn:normalised-rayleigh-quotient}
    R_\LL(g) =  \frac{\langle g, \LL g \rangle}{\langle g, g \rangle}
            = \frac{\sum_{u \sim v}A_{uv} \left(\frac{1}{\sqrt{d_u}} g(u) - \frac{1}{\sqrt{d_v}} g(v)\right)^2}{\sum_u g(u)^2}.
\end{equation}

When normalising signals by $D^{1/2}$~\cite{chung1997spectral} so that
\(R_\LL(D^{1/2}\mathbf{1})=0\), where $\mathbf{1}\in \RR^{n}$ is the vector of ones, we  obtain
\begin{equation}
\label{eqn:rayleigh-quotient-premultiplied}
R_\LL(D^{1/2}g) = \frac{\langle D^{1/2}g, \mathcal{L} D^{1/2}g\rangle}{\langle D^{1/2}g, D^{1/2}g\rangle}
= \frac{\langle g, L g\rangle}{\langle g, D g\rangle}
= \frac{
    \sum_{u \sim v}{A_{uv}
    (g(u) - g(v))^2}
    }{
    \sum_{u}{g(u)^2 d_u}
    }.
\end{equation}
The graph mean \(\mu_G(g)\) of a signal \(g\) is defined as
\begin{equation}
    \mu_G (g) = \frac{1}{\sum_{u\in V}d_u}\sum_{v\in V}{g(v)d_v}\label{eq:graph_mean}
\end{equation}
and the graph variance of $g$ is
\(\mathrm{Var}_G(g)=\sum_{v\in V}d_v\left(g(v)-\mu_G(g)\right)^2\)~\cite{He2005}.

\begin{definition}
If we re-centre the graph signal $g$ by setting $\tilde{g}(v)=g(v)-\mu_G(g)$, then the \emph{Laplacian score} of $g$ (in the sense of~\cite{govekClusteringindependentAnalysisGenomic2019}) 
is defined as
\begin{equation}
\label{eqn:LS}
LS(g)=R_\LL\left(D^{1/2}\tilde{g}\right)=\frac{\sum_{u \sim v} A_{uv}\left(g(u)-g(v)\right)^2}{\mathrm{Var}_G(g)}.
\end{equation}
\end{definition}

The Rayleigh quotient and Laplacian score measure consistency
of the graph signal with the underlying graph structure.
Small scores correspond to signals which exhibit variation consistent with the local graph structures; larger scores correspond to signals inconsistent with the local graph structures.
While the Rayleigh quotient is zero for constant signals (i.e., a perfect score),
the Laplacian score is undefined for constant signals and is high for
near-constant signals~\cite{chung1997spectral, He2005}.

{
Using the Laplacian matrix as a measure of consistency
of a graph signal
is directly related to Laplacian eigenmaps~\cite{belkin2003laplacian}. Laplacian eigenmaps construct an optimal (cf.~\cite{belkin2003laplacian}, Section 3.1) embedding of graph signals for
dimensionality reduction by finding the smallest eigenvalues of the generalised eigenvalue problem $Lg=\lambda Dg$. A signal $f$ solves this problem with eigenvalue $\lambda$ 
if $D^{1/2}$ is an eigenvector of $\mathcal{L}$ with eigenvalue $\lambda$. As $\mathcal{L}$ permits an orthonormal 
eigendecomposition, we can write $D^{1/2}\tilde{g}=D^{1/2}\sum_ia_i\tilde{g}_i$ where $D^{1/2}\tilde{g}_i$ is an eigenvector of $\mathcal{L}$ with eigenvalue $\lambda_i$. Then
$$LS(g)=\frac{\sum_i\lambda_ia_i^2}{\sum_ia_i^2}.$$
This score is small 
when $\tilde{g}$ aligns well with the Laplacian eigenmap embedding, i.e., when the squared coefficients $a_i^2$ corresponding to large eigenvalues are small.
}

\subsection{Eigenscores}\label{section:MM_eig}
The Rayleigh quotient and Laplacian score order graph signals by coherence with the underlying graph.
We remark that this ordering only considers consistency at a single scale. 
In order to obtain a finer-grained, multiscale understanding of graph signals, we consider their alignment with different coherent structures on multiple scales in the graph. To explain how we can do this, we first recall the spectrum of the Laplacian.

As both the Laplacian \(L\) and the normalised Laplacian \(\LL\)
are symmetric and positive semi-definite, 
their eigenvalues are real and non-negative~\cite{chung1997spectral}.
For the normalised Laplacian \(\LL\), 
write the orthonormal eigenbasis as
\(\{e_0, \ldots,  e_{n-1}\}\)
with corresponding eigenvalues
\(0=\lambda_0 \leq \lambda_1 \leq \cdots \leq \lambda_{n-1}\).

Given a graph signal \(g\), 
\(D^{1/2}g = \sum_{i=0}^{n-1} g_i e_i\)
where \(g_i = \langle D^{1/2}g, e_i \rangle\).
Writing \mbox{Equation~\eqref{eqn:rayleigh-quotient-premultiplied}} in this eigenbasis gives:
\begin{align}
    R_\LL(D^{1/2}g) &= 
    \frac{\langle\sum_i g_i e_i, \sum_j \lambda_j g_je_j \rangle}
    {\langle \sum_i g_i e_i, \sum_j g_j e_j \rangle}\notag \\
    &= \frac{\sum_i \lambda_i g_i^2}{\sum_i g_i^2}\notag\\
    &= \sum_i \lambda_i \left(\frac{g_i}{\|D^{1/2}g\|}\right)^2.\label{eq:eigenbasis}
\end{align}

Given the expression of the eigenbasis in Equation~\eqref{eq:eigenbasis}, we now consider individual contributions to the Rayleigh quotient separately, proposing the following definition.

\begin{definition}[Eigenscore]
Given a graph signal \(g: G \to \RR\) {and $e_i$ as the \(i\)th eigenvector of the normalised Laplacian,} 
we define the  { \(i\)th eigenscore \(\eig_i\)} by
\begin{equation}
\eig_i(g) = \frac{\langle D^{1/2}g, e_i \rangle}{\| D^{1/2}g \|}.
\end{equation}
{Given that \(g_i = \langle D^{1/2}g, e_i \rangle\),} it follows that \(R_\LL(D^{1/2}g) = \sum_i \lambda_i \eig_i(g)^2\).
\end{definition}

We can view the \(i\)th eigenscore of a graph signal
as the contribution from the \(i\)th eigenvector direction to its Rayleigh quotient.
It can also be viewed as the cosine of the angle between \(D^{1/2}g\)
and the \(i\)th eigenvector. 
Thus, a large positive value for \(\eig_i(g)\) indicates strong alignment
of the graph signal with the \(i\)th eigenvector of \(\LL\),
and a large negative value indicates strong anti-alignment (i.e., alignment with minus the eigenvector).

The ordering of the eigenvalues by magnitude
explains the multiscale nature of the eigenscore.
Expressing a graph signal in terms of Laplacian eigenvector contributions can be viewed as expanding in a frequency basis. Here, ordering the eigenvectors according to increasing eigenvalue corresponds to considering waves of increasing frequency. Expressing a signal in this basis can be viewed as the graph analogue of a Fourier transform. 
In general, computing the full eigendecomposition
is expensive; however, algorithms exist for computing
the first few dominant eigenvectors of a symmetric sparse matrix~\cite{calvetti1994implicitly}.

\subsubsection{The 0th Eigenscore}

Set $D^{1/2}\mathbf{1}$ to be the 0th eigenvector in our eigenbasis. Then
$$
\eig_0(g) = \frac{\langle D^{1/2}g,  D^{1/2} \mathbf{1}\rangle}{\| D^{1/2}g\| \|D^{1/2} \mathbf{1}\|}
=
\| D^{1/2} \mathbf{1} \| 
\, \, \mu_G\left(\frac{g}{\| D^{1/2}g\|} \right),
$$
where $\mu_G$ is the graph mean, as defined in Equation \eqref{eq:graph_mean}.

\subsubsection{Eigenscores to Visualise Graph Signals}
Projecting gene signals onto the eigenspace spanned by low-frequency eigenscores
allows us to visualise gene space and identify meaningful signals (see Figure~\ref{fig:eig_explain}). Noisy signals are mapped close to zero, and interesting signals lie on the periphery in such an eigenspace plot.
Constructing such an embedding using Laplacian eigenvectors is reminiscent of Laplacian eigenmaps~\cite{belkin2003laplacian}. However, in~\cite{belkin2003laplacian}, Laplacian eigenvectors are used to construct an embedding of the nodes of the graph, whereas we embed signals on the graph.

\subsection{Multiscale Laplacian Score}\label{section:MM_MLS}

The Laplacian score (Equation~\eqref{eqn:LS}) 
considers the change in signal along single edges in the graph.
We propose the multiscale Laplacian score (MLS), which
relies on random walks
to measure the consistency of a signal across local graph neighbourhoods of continuously increasing size. This unsupervised approach provides a multiscale ranking of signal coherence with the graph.
We can determine a finite number of scales at which the random walker admits a Markov stable partition~\cite{delvenne2010stability,lambiotte2014random}, and we pair this pipeline with the MLS.

\subsubsection{Random Walks on Graphs}

Random walks on graphs are stochastic processes that can model a range of phenomena, including diffusion on graphs~\cite{masuda2017random}.
For any graph $G$ with adjacency matrix $A$, the evolution of a continuous-time Markov process is governed by the Kolmogorov differential equation:
\begin{equation}
\Dot{\mathbf{p}}=-\mathbf{p}L^\mathrm{rw},\label{eq:de}
\end{equation}
where {$\mathbf{p}$} is a time-dependent node vector and $\mathbf{p}_v (t)$ gives the probability of a random walker being on node $v$ at time $t$.
In this Markov process, a random walker jumps to adjacent nodes (with probability proportional to the respective edge weight) after a period of time drawn from an $\mathrm{Exp}(1)$ random variable.
The stationary distribution $\pi\in \RR^{n}$ is the unique left eigenvector of $L^\mathrm{rw}$ with eigenvalue 0 whose entries sum to 1. The solution to Equation \eqref{eq:de} is 
$\mathbf{p}(t)=\mathbf{p}(0)\exp(-tL^\mathrm{rw})$
and $\pi=\lim_{t\to\infty}\mathbf{p}(t)$.

\begin{remark}
{
{The} heat kernel can be used in conjunction with the graph Laplacian to model heat flow on a graph, which is viewed as a discrete approximation to a Riemannian manifold (see~\cite{belkin2003laplacian} Section~3.3 and references therein). From this viewpoint, with signals representing heat, the Rayleigh quotient quantifies relative differences in heat across the manifold for an infinitesimal time step.} 
\end{remark}

\subsubsection{Community Detection}

Community detection in networks is concerned with finding groups of nodes that are more tightly connected to each other than to the rest of the network. Some of the best known community detection algorithms, such as modularity optimisation~\cite{porter2009communities}, exploit combinatorial properties of the graph. The communities found by optimising modularity are {\it {dense}}; i.e., there are many more edges between nodes in the group than with the rest of the network.

Community detection has extended from the notion of dense connections defining a community to also include connectivity via random walks. Markov stability, a dynamical approach for community detection, relies on random walks to detect stable graph partitions $V=C_1\sqcup C_2\sqcup\dots\sqcup C_k$ at multiple resolutions~\cite{delvenne2010stability,lambiotte2014random}. 
We call each $C_i$ a community and assume that it is non-empty. Moreover, we denote the community to which  node $v$ belongs as $c_v$ and assume that all subgraphs induced by $C_i$ are connected. Two partitions are 
considered identical if one can be obtained from the other by permuting the labels $1,...,k$.

\begin{definition}
Let $\{C_i\}_{i=1,...,k}$ be a partition of the graph $G$ into communities.
If $M=D^{-1}A$ is the random walk transition matrix with stationary distribution $\pi$, then the 
 {continuous Markov stability} of the partition at 
time is
$$r_\mathrm{cont}(\{C_i\},t)=\sum_{u,v\in V}\pi_u\left(P(t)_{uv}-\pi_v\right)\delta(c_u,c_v),$$
where $\delta$ is the Kronecker delta and $P(t):=\exp(-tL^\mathrm{rw})$ is the continuous time transition matrix~\cite{Delvenne2013}.
\end{definition}

The Markov stability of a graph partition at time $t$ is the probability of a random walker remaining in its initial community after walking for time $t$ minus the probability that two independent random walkers are in the same community at time $t$. All walkers are assumed to be in the stationary distribution.
The Markov stability of a partition $\{C_i\}$ at time $t$ takes values in the range $(-1/2, 1]$. High values indicate that a random walker tends to be trapped in one of the groups, which is what we expect in the presence of communities.
For each value of $t$, coherent community structures on a graph can be found by maximising Markov stability 
using the Louvain method~\cite{Blondel2008}, which is a successful algorithm for finding community structures at different scales in applications~\cite{bacik2016flow, Beguerisse2013, liu2020graph}. {A 
bottleneck (for graphs larger than those considered 
in this study) is the computation of the matrix exponential for the Markov stability, which requires an eigenvalue decomposition of $L^\mathrm{rw}$. To speed up computations in such regimes, one can use a linear approximation of $r_\mathrm{cont}$ for small $t$ (see~\cite{Delvenne2013} Section 3.3).}

\begin{figure}[h]
 
    \includegraphics[width=\textwidth]{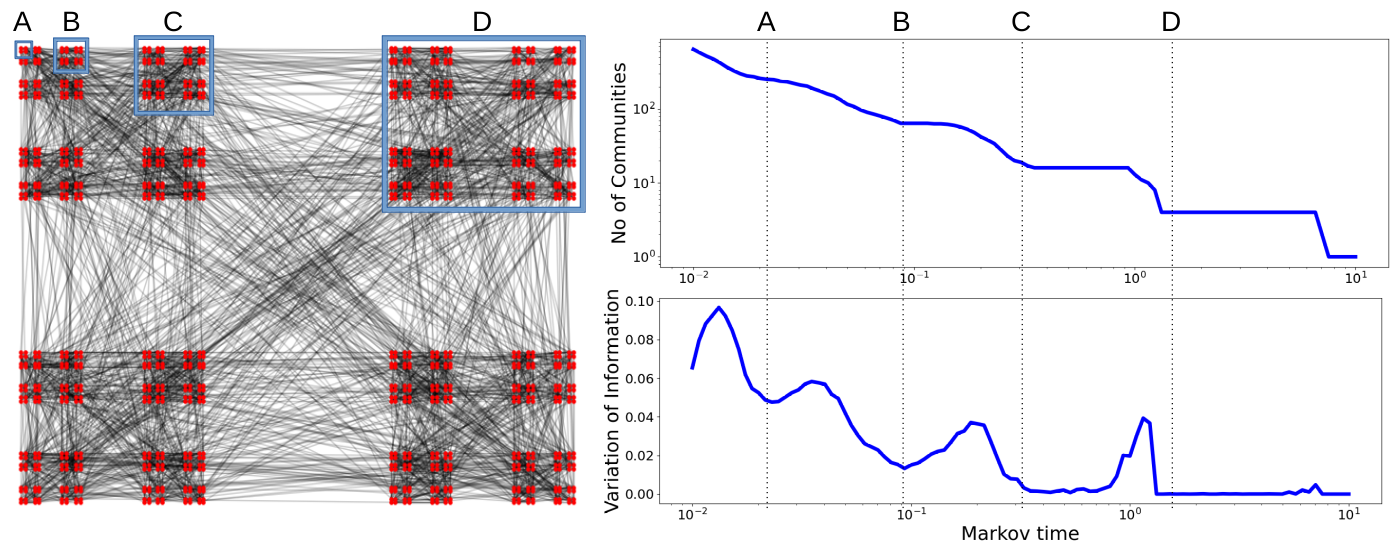}
    \caption{The graph on the \textbf{left} displays community structures at four different scales, exemplified by the groups A, B, C and D.
    When computing the mean pairwise variation of information (\textbf{right}) as a function of scale (Markov time), we find local minima corresponding to resolutions A (256 communities), B (64 communities), C (16 communities) and D (4 communities). Figure inspired \mbox{by~\cite{StabilityEg}}.}
    \label{fig:vi_eg}
\end{figure}

The choice of values of $t$ to use for finding community structures via the maximisation of Markov stability depends on the graph $G$. For example, a complete graph will only have one sensible community structure (containing a single community) which will be detected at a relatively large $t$, while many real-world networks exhibit community structures at a variety of scales $t$. 
The partitions at different $t$ obtained from this optimisation are assessed using the mean pairwise \emph{{variation} of information (VI)}, which
tests the consistency and robustness of partitions~\cite{meilua2007comparing}:

{
\begin{definition}
Let $\{C_i\}_{1\leq i\leq k}$ and $\{C'_j\}_{1\leq j\leq k'}$ be two partitions of a graph with $N$ nodes. Then, their variation of information is defined to be
$$\mathrm{VI}\left(\{C_i\}, \{C'_j\}\right)=2H\left(\{C_i\}, \{C'_j\}\right)-H(\{C_i\})-H(\{C'_j\}),$$
where
$$H\left(\{C_i\}, \{C'_j\}\right)=-\sum_{i,j}\frac{|C_i\cap C'_j|}{N}\log_2\left(\frac{|C_i\cap C'_j|}{N}\right),\; H\left(\{C_i\}\right)=-\sum_{i=1}^k\frac{|C_i|}{N}\log_2 \left(\frac{|C_i|}{N}\right).$$
\end{definition}
The VI provides a measure of similarity between partitions, which can be viewed as a function of $t$. At resolutions of $t$ for which there is an obvious community structure, the VI is relatively small and takes a local minimum---a plateau of such a minimum suggests the stable partition of communities.}  This behaviour is explained in~\cite{Schaub2012} and illustrated in Figure \ref{fig:vi_eg}. 
{In this work, we chose values of $t$ for which 
VI attains a local minimum. Each graph may have different stable communities and, therefore, the selected value of $t$ would be chosen based on 
the minima of that graph.}

\subsubsection{Signal Scores at Multiple Resolutions}

We can reinterpret the Laplacian score (Equation~\eqref{eqn:LS}) in
terms of the random walk Laplacian (Equation \eqref{eq:rw_laplacian}):
\begin{equation}
    LS(g)=\frac{\left\langle D^{1/2}\tilde{g}, D^{1/2}L^\mathrm{rw} \tilde{g}\right\rangle}{\left\langle D^{1/2}\tilde{g}, D^{1/2}\tilde{g}\right\rangle} = \frac{\sum_{u, v\in V} d_u(D^{-1}A)_{uv}\left(g(u)-g(v)\right)^2}{2\cdot\mathrm{Var}_G(g)}. \label{eq:LS-rw}
\end{equation}
Thus, the Laplacian score of a signal $g$
is the expected squared difference in the signal $g$ that is observed when a random walker at stationary distribution takes exactly one step following transition matrix $D^{-1}A$, which is divided by twice the graph variance.
By extending the Laplacian score from a single random step to a random walk for time $t$,
we arrive at our definition for the multiscale Laplacian score:
\begin{definition}
Let $G=(V,E)$ be a graph with adjacancy matrix $A$, $g:V\to\mathbb{R}$ be a signal on $G$ and $t\in\mathbb{R}_{\geq 0}$. The  {multiscale Laplacian score} of $g$ at resolution $t$ is defined as
$$
MLS(g, t)=\frac{\left\langle D^{1/2}\tilde{g}, D^{1/2}(I-P(t)) \tilde{g}\right\rangle}{\left\langle D^{1/2}\tilde{g}, D^{1/2}\tilde{g}\right\rangle}=\frac{\sum_{u, v\in V}d_uP(t)_{uv}\left(g(u)-g(v)\right)^2}{2\cdot\mathrm{Var}_G(g)},
$$
where we use the identity $d_uP(t)_{uv}=d_vP(t)_{vu}$ for all $t\in\RR_{\geq0}$ and all $u,v\in V$.
\end{definition}

If the expected change in a signal $g$, which a continuous-time random walker is exposed to, 
after time \(t\) is small, then the $\mathrm{MLS}(g,t)$ is small.
In such a case, we say that the signal $g$ is consistent with the graph structure of $G$ at resolution $t$. The MLS extends the Laplacian score (Equation~\eqref{eqn:LS})~\cite{govekClusteringindependentAnalysisGenomic2019} by performing a consistency analysis at multiple resolutions. Analysing multiple resolutions, ranging from local to global structures, is useful for studying graphs $G$ paired with signals that are consistent at multiple resolutions. 

\subsubsection{MLS Analysis Pipeline}

\begin{figure}[h]

    \includegraphics[width=0.7\textwidth]{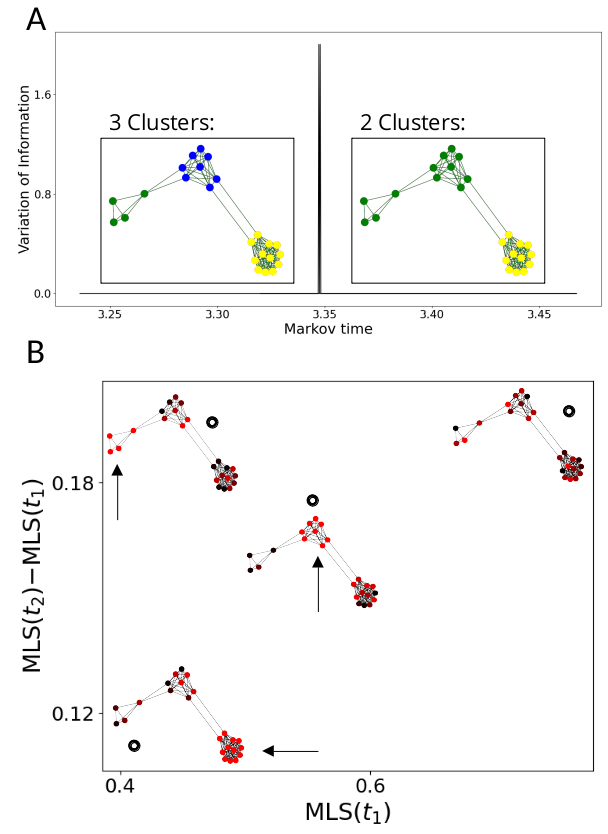}
    \caption{We construct a graph with three communities, all of different sizes. (\textbf{A}) The VI (on y-axis, VI is 0 except for a brief spike around $t=3.35$) identifies resolutions $t_1$, at which all three communities are identified, and $t_2,$ at which two communities are identified 
    (note that due to the simplicity of the graph, 
    there are intervals of local minima instead of points; we pick $t_1$ before the spike and $t_2$ after). 
    In (\textbf{B}), we calculate the MLS at $t_1$ and $t_2$ (given by black circles) of three signals that are equal to 1 on one of the $t_1$-communities (constant part of the signal is highlighted by arrows) and uniformly random elsewhere, and one completely random signal. The signal that is constant on the largest cluster (\textbf{bottom left}) is identified as highly consistent at both times. The random signal (\textbf{top right}) is identified as inconsistent at both times. Conversely, the signal constant on the smallest community (\textbf{top left}) has a high MLS at $t_2$ relative to the MLS at $t_1$, separating it from the signal constant on the community of intermediate size (\textbf{centre}).
}
    \label{fig:mls_eg}
\end{figure}

In the MLS analysis pipeline, we partition a given graph $G$ into communities at 100 Markov times using the Louvain algorithm~\cite{Blondel2008}. 
We then select a small set of Markov times at which the VI attains local minima. Next, we calculate the MLS at each of these resolutions and for each signal on the graph. 
We can then compare the MLS at different Markov times to identify gene signals particularly consistent with a given topological structure at a given resolution. For example, a small MLS at an earlier Markov time (compared to the mean behaviour of all signals) is more consistent with structures at that resolution (see Figure \ref{fig:mls_eg}).

\subsection{Persistent Rayleigh Quotient}\label{section:MM_PRQ}

Given a graph $G$ and signals $g$:$V \rightarrow \RR$, we may have additional information associated to each node of \(G\)
that we would like to use to further inform our analysis.
In single-cell data, for example, this could be the developmental
time of each observed cell, which associates a real
value to each node of the graph \(G\).

\begin{definition}[Filtered graph]
A filtration of a graph \(G\)
is a integer-valued function {\(f : V \to \mathbb{Z}\)}
on the nodes of \(G\).
For \(i \in \mathbb{Z}\)
the sub-level set \(\alpha(i)\)
of \(f\) at \(i\) is the set
\[
\alpha(i) = \{v \in V: f(v) \leq i\}
,\]
the nodes of \(G\) have a filtration value not greater than \(i\). 
The induced subgraph \(G[\alpha(t)]\)
is the subgraph of \(G\) with nodes \(\alpha(i)\)
and every edge in \(G\) that has both
endpoints in \(\alpha(i)\).
Then, the filtration \(f\) gives a sequence
of induced subgraphs of \(G\)
\[
G[\alpha(i_0)] \leq G[\alpha(t_1)] \leq \cdots \leq G[\alpha(i_n)] \leq G
\]
for each increasing sequence \((i_k)_{k=0}^n\) of real numbers.
\end{definition}

Topological data analysis studies the evolution of topological invariants across
filtered graphs \(G[\alpha(i)]\).
The most common tool is persistent homology~\cite{ghrist2008barcodes},
which computes how invariants, such as connected components
in \(G[\alpha(i)]\), persist in the larger graph \(G[\alpha(j)]\).
Persistent homology is limited to studying the structure of the filtered graph itself. To analyse the signals on the sequence of subgraphs, we first recall the persistent Laplacian and then introduce the persistent Rayleigh quotient.

\subsubsection{Persistent Laplacian}

Given a subset \(\alpha \subseteq V\), one
can reduce the Laplacian of \(G\)
to a Laplacian on the nodes \(\alpha\)
by a method known as Kron reduction~\cite{Dorfler2013}.
Briefly, Kron reduction removes the nodes in \(V\setminus{\alpha}\)
and adds weighted edges that preserve the geometric
structure between the nodes \(\alpha\) in \(G\).
For example, in network circuit theory~\cite{Dorfler2013},
Kron reduction creates a simpler representation
of a circuit whilst
preserving resistances.
Memoli, Wang and collaborators extended this method to higher-order 
graphs (i.e., simplicial complexes) 
\cite{memoliPersistentLaplaciansProperties2021, wangPersistentSpectralGraph2019a}
and showed that the Kron reduced Laplacian is the 0-degree persistent Laplacian.
There is a direct relationship
between persistent homology
and the Kron reduction/persistent Laplacian:
the nullity of the reduced Laplacian is
exactly the persistent Betti number
of \(G[\alpha] \subseteq G\)~\cite{memoliPersistentLaplaciansProperties2021}.
For graphs, this persistent Betti number is
the number of connected components
of \(G[\alpha]\) that remain
disconnected in \(G\).

For subsets \(\alpha, \beta \subseteq V\)
let \(L[\alpha, \beta]\)
be the submatrix of \(L\)
with rows indexed by \(\alpha\) and columns indexed by \(\beta\).
Under an appropriate reordering
of the node labels, the Laplacian \(L\) has block form
\[
L = 
\begin{bmatrix}
L[\alpha, \alpha] & L[\alpha, \ac] \\
L[\ac, \alpha] & L[\ac, \ac]
\end{bmatrix},
\]
where \(\ac = V \setminus \alpha\)
is the complement of \(\alpha\) in \(V\).

\begin{definition}[Kron reduction~\cite{Dorfler2013}/Persistent Laplacian~\cite{memoliPersistentLaplaciansProperties2021}]
\label{def:kron_reduction}

The Kron reduction
(or 0-degree persistent Laplacian)
of \(L\) with respect to \(\alpha\)
is the matrix
\[
L_{\alpha} 
= L[\alpha, \alpha]
- L[\alpha, \ac] L[\ac, \ac]^{-1} L[\ac, \alpha],
\]
which is also known as the Schur complement
\(L/L[\ac, \ac]\).
We analogously define \(\LL_\alpha\)
for the normalised Laplacian \(\LL\).
\end{definition}

The Kron reduction \(L_\alpha\) of $L$
arises from performing Gaussian elimination on \(L\)
to remove blocks \(L[\alpha, \ac]\) and \(L[\ac, \alpha]\):
\[
\begin{bmatrix}
L[\alpha, \alpha] & L[\alpha, \ac] \\
L[\ac, \alpha] & L[\ac, \ac]
\end{bmatrix}
\rightsquigarrow
\begin{bmatrix}
L[\alpha, \alpha]
- L[\alpha, \ac] L[\ac, \ac]^{-1} L[\ac, \alpha]
& 0 \\
0 & L[\ac, \ac]
\end{bmatrix}
.
\]

\begin{lemma}[lemma 2.6 in~\cite{Dorfler2013}]
In definition \ref{def:kron_reduction}, the following hold:
\begin{enumerate}
    \item \(L_\alpha\) is well-defined as \(L[\ac, \ac]\) is invertible.
    \item \(L_\alpha\) is symmetric.
    \item \(L_\alpha\mathbf{1} = \mathbf{0}\), where \(\mathbf{1}\) is the column vector of ones.
\end{enumerate}
\end{lemma}
Hence, \(L_\alpha\) is a Laplacian matrix in the sense
that there exists a weighted graph with nodes \(\alpha\)
and Laplacian equal to \(L_\alpha\).

Suppose we have a filtration \(f\) on the nodes of the graph \(G\).
Then
for \(i, j \in \mathbb{Z}\) with \(i \leq j\), define
\[
L_i^j =\left(L^{\alpha(j)}\right)_{\alpha(i)}
\]
the \((i,j)\)-persistent Laplacian,
where \(L^{\alpha(j)}\)
is the Laplacian of the graph \(G[\alpha(j)]\). 
Again, $\LL_i^j$ is defined analogously.

\begin{definition} [Persistent Rayleigh quotient]
For a graph \(G\) with filtration \(f\) 
and \(i, j \in \mathbb{Z}\) with \(i \leq j\),
the
 {persistent Rayleigh quotient}
of a signal \(g: G \to \RR\) is

\[
\PRQ(i, j)(g) = R_{L_i^j}(g) 
= \frac{\langle g , L_i^jg \rangle}{\langle g , g \rangle},
\]
which is the Rayleigh quotient (as in Equation~\eqref{eqn:rayleigh-quotient}) using the \((i, j)\)-persistent Laplacian.

We further define the normalised persistent Rayleigh quotient to be
\[
\NPRQ(i, j)(g) = R_{\LL_i^j}((D_i^j)^{1/2}g) 
= \frac{\langle g , L_i^jg \rangle}{\langle g , D_i^j g \rangle},
\]
which is the Rayleigh quotient (as in Equation~\eqref{eqn:rayleigh-quotient-premultiplied}) using the normalised version of the \((i,j)\)-persistent Laplacian
on the normalisation of the signal.
Here, \(D_i^j\) is the degree matrix of
the graph corresponding to \(L_i^j\).
When applying \(L_i^j\) and \(D_i^j\) to \(g\),
we implicitly restrict \(g\) to the nodes \(\alpha(i)\). 
\end{definition}
\subsubsection{Application to Cell Bifurcation}
\begin{figure}[h]
 
    \includegraphics[width=0.7\textwidth]{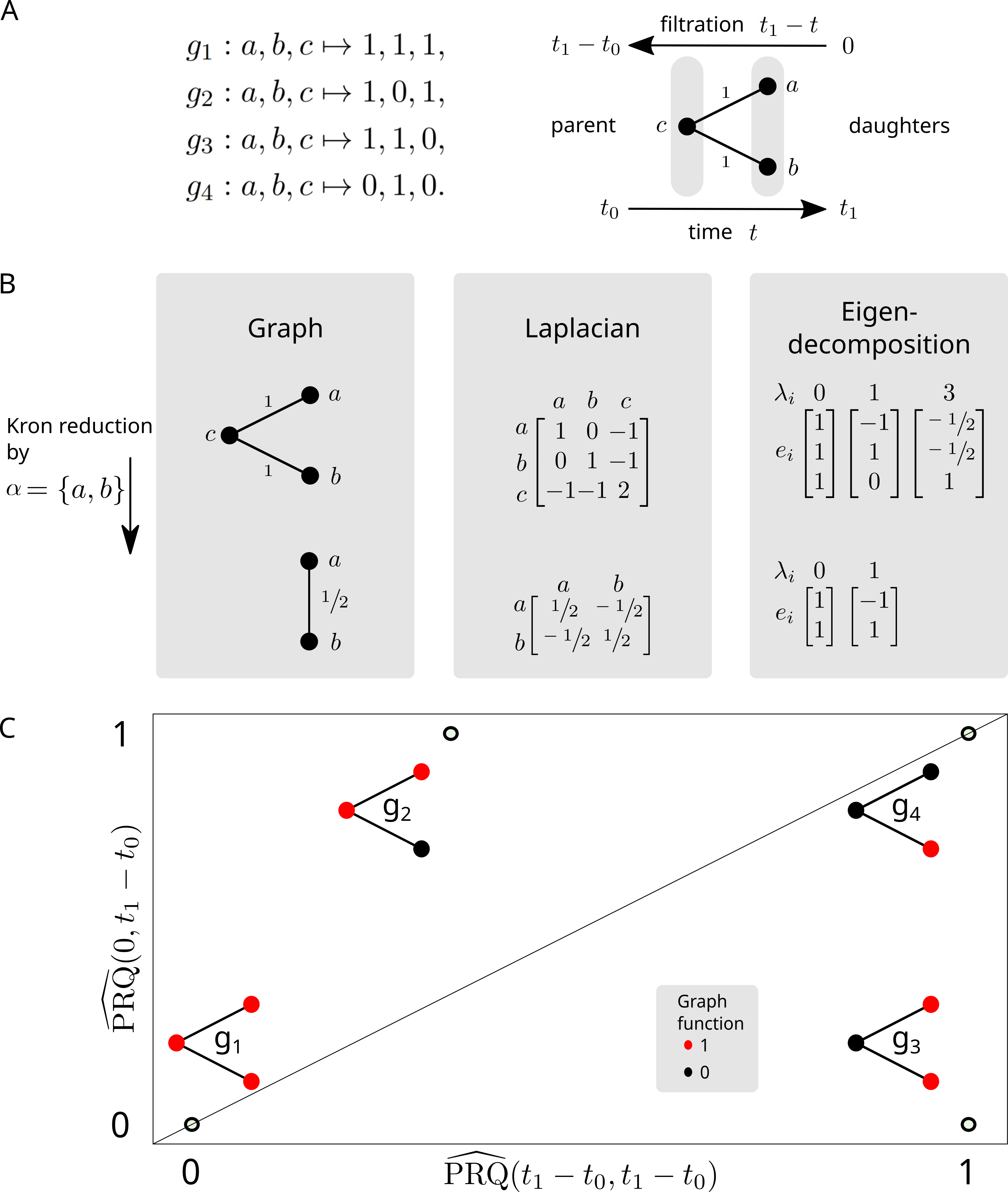}
    \caption{
    The persistent Rayleigh quotient for cell differentiation.
    (\textbf{A}) {(\textbf{left}) Signals (genes) on the graph that we aim to differentiate.}
    (\textbf{right}) The model for the bifurcating differentiation process.
    (\textbf{B}) The effects on the graph and graph Laplacian
    after applying the Kron reduction process to the daughter cells.
    (\textbf{C}) The normalised Rayleigh quotients of (\emph{x}-axis) full Laplacian \(L_{t_1 - t_0}^{t_1 - t_0}\)
    and (\emph{y}-axis) persistent Laplacian \(L_0^{t_1 - t_0}\)
    for binary functions on the graph representing high and low gene expression
    of a particular gene.
    {The persistent Rayleigh quotient separates these genes
    based on relevance to the bifurcation:
    \(g_1\) is expressed in all cell types,
    \(g_2\) is expressed in the parent and one daughter cell type,
    \(g_3\) is expressed only in both daughter cell types,
    \(g_4\) is expressed only in one daughter cell type.}
    }
    \label{fig:pl_explain}
\end{figure}

We demonstrate the persistent Rayleigh quotient on a toy bifurcation model \(G\) where
\(V = \{a, b, c\}\)
and \(E = \{(a, c), (b, c)\}\)
(Figure  \ref{fig:pl_explain}).
We consider the graph signals
\begin{align*}
    & g_1 : a, b, c \mapsto 1, 1, 1 , \\
    & g_2 : a, b, c \mapsto 1, 0, 1 , \\
    & g_3 : a, b, c \mapsto 1, 1, 0 , \\
    & g_4 : a, b, c \mapsto 0, 1, 0 .
\end{align*}
Suppose that this graph represents a biological system: node \(c\) represents a parent cell type at developmental time \(t_0\)
and nodes \(a\) and \(b\) are daughter cell types at developmental time \(t_1\).
If we filter the graph \(G\) by time,
then we only ever have one connected component.
Thus, we filter in `reverse time' by setting the filtration to be
\(t_\text{max} - t\). 
Explicitly, we define a filtration \(f\) by \(f(a) = f(b) = 0\) and \(f(c) = t_1 - t_0\).
Now, \(G[\alpha(0)]\) has two connected components which merge into a single component in
\(G = G[\alpha(t_1 - t_0)]\).

When we perform the Kron reduction of \(L=
L^{t_1 - t_0}\) with respect to \(\alpha(0)\) to obtain the
Laplacian \(L_{0}^{t_1 - t_0}\), the graph associated to this Laplacian has just \(a\) and \(b\) as
nodes and a single \(1/2\)-weight edge between them (Figure \ref{fig:pl_explain}B).
This graph still has one
connected component,
but the connection is weaker.
In the language of persistent homology, this corresponds to
two \(H_0\)-bars: one is born at filtration value \(0\) and dies before value \(t_1 - t_0\), and the other is born at value \(0\) and persists infinitely.

Comparing the usual normalised Rayleigh quotient, corresponding to \(\NPRQ(t_1 - t_0, t_1 - t_0)\), to the normalised persistent Rayleigh quotient
\(\NPRQ(0, t_1 - t_0)\) separates the binary graph functions \(g_i\) on \(G\) (Figure \ref{fig:pl_explain}C).
In the context of single-cell differentiation data, graph signals correspond to genes.
Gene \(g_2\) lies
above the diagonal in Figure \ref{fig:pl_explain}C and is highly expressed
in the parent cell type and only one of the daughter cell types.
{
Such behaviour indicates that a gene, or its transcriptional regulators, may
be involved in cell differentiation towards a particular lineage or fate.
}
Similarly, gene \(g_3\), which lies below the diagonal, is expressed in both daughter cell types
but not the parent cell type, and, as such, it may
correspond to 
{a class of genes which 
represents markers of differentiation shared across cell fates.}
Genes corresponding to \(g_1\) are
expressed at a constant rate, 
representing possible `house-keeper' genes and, thus, have
zero (or close to zero) persistent Rayleigh quotients.
Finally, gene \(g_4\) is only expressed in a single daughter cell type and, as such, {represents markers of differentiation  
to a particular cell fate}. Gene \(g_4\) lies along the diagonal.

\subsection{Data Sets}\label{sec:data}

{We apply the proposed methods on three different experimental scRNA-seq data sets. The first is an scRNA-seq data set of 2700 human peripheral blood mononuclear cells (PBMC)~\cite{10XPBMC}, which is used in Seurat tutorials ~\cite{seuratPBMC,seuratVST}. The cell types found in the PBMC data set are lymphocytes (T cells, NK cells, B cells), monocytes, and dendritic cells. 
This data set also contains a small group of \textit{PPBP}+ cells, which have been variously identified as platelets or their progenitors, megakaryocytes
\cite{seuratPBMC,seuratVST,scanpy}. For simplicity, we will refer to them as platelets in our discussion, following the Seurat tutorial nomenclature.}

{
The second data set is scRNA-seq of 24,911 human T cells infiltrating lung tumours and adjacent normal tissue~\cite{lambrechts2018phenotype} (previously analysed with the Laplacian score~\cite{govekClusteringindependentAnalysisGenomic2019}). While the majority of the cells are CD4+ and CD8+ T cells, some NK cell clusters are also present in this data set.
The aim of the T cell experimental data set was to determine how the cellular composition of the stromal region surrounding lung tumours differs from the stromal region of healthy lung tissue. The authors analysed scRNA-seq data to identify different stromal cell compositions (or signatures) that correlate with survival.}

{
The third data set is scRNA-seq of 447 mouse foetal liver cells at different stages of development ~\cite{yangSinglecellTranscriptomicAnalysis2017}. 
The aim of the mouse liver experimental data was to identify divergence in gene expression (and associated time-course) as progenitor cells in the liver (i.e., hepatoblasts) specialise or differentiate into two different types of mature cells, hepatocytes and cholangiocytes.
Cells in the mouse foetal liver data set were sampled on embryonic days 10, 11, 12, 13, 14, 15, and 17. }

\subsubsection{Preprocessing of PBMC and T Cell Data Sets}
We normalised the PBMC and T cell scRNA data sets using the variance stabilizing transform (VST)~\cite{hafemeister2019normalization} 
{as implemented by the function SCTransform from the \texttt{R}-library Seurat~\cite{SeuratV4}}. 
The VST returns the 3000 genes with the highest dispersion in each data set, and  
it is then reduced to its 30 principal components with the highest variance,
following the recommendation given in the manual of Seurat.
We then construct a $k$-nearest neighbour ($k$-nn) graph on cells for both data sets, using $k=15$, and weight the edges of these graphs according to the weights given by the dimension reduction algorithm UMAP~\cite{mcinnes2018umap}.
{We note that the method of eigenscores is particularly robust to changes in the value of $k$ as well as leaving out preprocessing with PCA (results not shown).}
We use cosine-dissimilarity for the PBMC data (following the Seurat tutorials) and Pearson correlation-distance for the T cell data (following~\cite{govekClusteringindependentAnalysisGenomic2019}). We sample 3000 cells at random between the PCA and $k$-nn graph steps in the T cell data set (following~\cite{govekClusteringindependentAnalysisGenomic2019}).

\begin{remark}
{
{The} UMAP visualisation approach uses a locally scaled Laplacian kernel to weight its edges. A Laplacian kernel is similar, but not identical, to a heat kernel (also known as Gaussian kernel) in its definition. We use the weighted graph generated by UMAP in all of our analyses.}
\end{remark}

\subsubsection{Preprocessing of Mouse Foetal Liver Cell Data Set}
As the mouse data set is substantially smaller than the other data sets,
we used a simpler preprocessing strategy.
The public data from~\cite{yangSinglecellTranscriptomicAnalysis2017} 
were provided in transcripts-per-million (TPM),
and we further applied a \(\log_{e}(x + 1)\) transform.
The 10,000 most highly varying genes were retained. {We tailored the value of k according to the number of cells in the experiment. For the larger experiments, we used the default $k=15$ whereas here, } the UMAP-weighted $k$-nn graph was built using \(k=3\) with the Euclidean metric. 
As in the original paper, we plot the resulting graph on the first two principal components \mbox{(Figure \ref{fig:mouse-data set}).}

\subsubsection{Previous Results on PBMC Data}
{
In the UMAP plot generated from the variance stabilised PBMC data (see Figure \ref{fig:pbmc_clusters}), five large populations (clusters) and two smaller ones are visible. 
By colouring the UMAP plot by marker gene expression, the VST vignette in~\cite{seuratVST} identifies that the large populations roughly correspond to CD4 T cells, CD8 T cells, NK cells, B cells and monocytes, while the small components contain platelets and dendritic cells.
The clustering algorithm applied to the data further subdivides these populations. By performing differential gene expression (DGE) analysis by use of a non-parametric Wilcoxon rank sum test~\cite{seuratDGE}, the VST vignette suggests that the CD4 and CD8 T cells can be split into three subpopulations.
The B cells and NK cells are decomposed into two subpopulations, each closely linked to a high expression of known marker genes. The top ten differentially expressed genes in the twelve resulting clusters are given in Table \ref{Table:pbmc_dge}.
}

\subsubsection{Previous Results on T Cell Data}
\begin{figure}[h]
 
	\includegraphics[width=\textwidth]{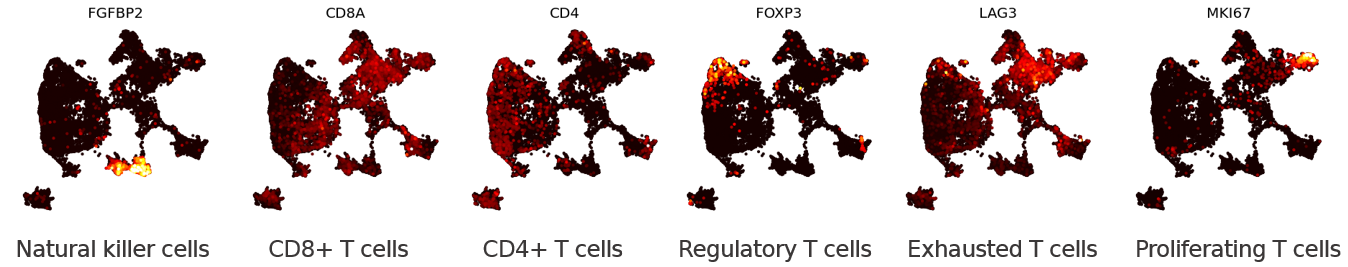}
	\caption{Lambrechts et al.~\cite{lambrechts2018phenotype} classified T cells into six sub-cell types based on marker genes. {
	To reduce overplotting and assist visualisation, points with non-zero expression were plotted on top for this figure and Figure \ref{fig:extra_genes}.}}
	\label{fig:tcell_clusters}
\end{figure}

{
Lambrechts et al.~\cite{lambrechts2018phenotype} identify several subclusters of T and natural killer (NK) cells based on clustering in t-SNE (nine clusters) and marker genes (six subgroups, see Figure \ref{fig:tcell_clusters}). There is only one visibly connected component in the original t-SNE plot, while our UMAP plot shows more structure. Govek et al.~\cite{govekClusteringindependentAnalysisGenomic2019} applied the combinatorial Laplacian score 
to identify a variety of genes, including \textit{HAVCR2}, \textit{RSAD2} and \textit{GZMK}, that are consistent with the topological structure of the data set but do not correspond to the clusters defined in~\cite{lambrechts2018phenotype}. 
Govek et al. also demonstrated on the T cell data that the discriminating power of the combinatorial Laplacian score (measured by the area under the receiver-operating characteristic curve) is comparable to that of conventional DGE methods and  superior to feature variance without taking topology into account
\cite{govekClusteringindependentAnalysisGenomic2019}.}

\subsubsection{Previous Results on Mouse Foetal Liver Cell Data Set}
Yang et al.~\cite{yangSinglecellTranscriptomicAnalysis2017}
selected 1761 heterogeneously expressed genes
that correlate with the first two principal components
of the data which were then clustered.
In a separate study on different {data}, Mu et al.~\cite{muEmbryonicLiverDevelopmental2020}
ranked genes that were differentially expressed in
hepatocyte development.

\subsection{Code Availability}
The code for computing eigenscores, the multiscale Laplacian score,
and the persistent Rayleigh quotient for signals on a graph is available here:

\href{https://github.com/osumray/multiscale-signal-selection-single-cell.git}{https://github.com/osumray/multiscale-signal-selection-single-cell.git} (accessed on 11 August 2022).

%% file: results.tex
\section{Results}\label{section:Results}

In this section, we apply the three multiscale methods outlined above
to three single-cell data sets. 
Eigenscores rank genes at different frequencies: high eigenscores identify dominant genes that align with the underlying cell similarity graph as the frequency increases.
The multiscale Laplacian score (MLS) identifies coherent genes as the distance traversed by a random walker on the cell similarity graph is scaled. Third, the persistent Rayleigh quotient identifies genes involved in bifurcation processes when additional temporal meta data are available.

\subsection{Eigenscores}\label{section:Results_eig}

Eigenscores test whether features such as genes, viewed as functions on nodes representing cells of a graph, align or anti-align with the Laplacian eigenvectors on the graph. 
They can be used to rank genes similarly to DGE but on different scales in the data, according to their coherence with the topology of the cell similarity graph. 
They can also be applied to explore gene expression by visualising genes in a gene space. 
By scoring genes for alignment with individual eigenvectors, as well as selecting relevant genes, we can often {shed light on} the biological processes in which these genes are involved.
Eigenscores are meaningful in data sets with clear community structures, e.g., the PBMC data set. Eigenscores are also meaningful in data sets that cannot be decomposed into distinct clusters but rather have a continuous structure, e.g., the T cell data set. In either case, the eigenscores do not rely on predefined clusters; they scan through the data in an unsupervised way.

\clearpage

\subsubsection{The Geometry of PBMC Genes via Eigenscores}
\begin{figure}[h]
 
	\includegraphics[width=\textwidth]{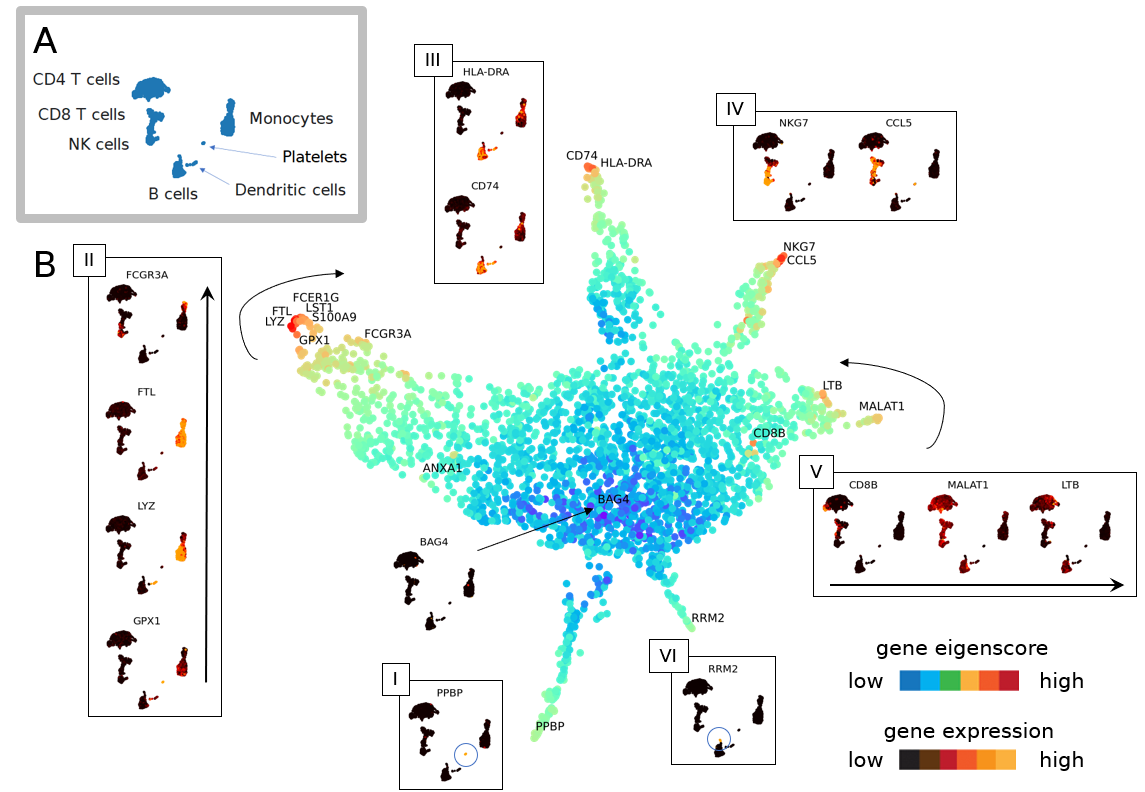}
	\caption{Geometry of cell space and gene space. (\textbf{A}) Cell types in PBMC data~\cite{10XPBMC}. 
	(\textbf{B}) UMAP of genes set in eigenscore space for eigenvectors 1--16. Genes (dots) are colour-coded for the logarithm of the norm of the vector in 16-dimensional eigenscore space. 
	Genes with similar expression patterns in the PBMC single-cell data~\cite{10XPBMC} plot close together in eigenscore space, and expression patterns vary continuously as we move through this space. The outward branches I--VI correspond to genes that
	are expressed highly on specific groups of cells. 
	}
	\label{fig:eigenvector_PBMC}
\end{figure}

We compute eigenscores of the top genes in the PBMC data set~\cite{10XPBMC} (see Figure \ref{Table:pbmc_eig}). 
While many selected genes {
overlap with cluster marker genes identified by differential gene expression (DGE) via Seurat's default implementation of the non-parametric Wilcoxon rank sum test~\cite{seuratDGE}}, validating eigenscores, this method identifies 26 additional genes (see Table \ref{Table:pbmc_dge}). 
To interpret the genes, we compare to 12 finer cell subtypes or seven
previously determined broader cell subtypes~\cite{seuratPBMC} (see Figures~\ref{fig:eigenvector_PBMC}A and \ref{fig:pbmc_clusters}). Eigenscores select genes that {are enriched in} broader cell types (e.g., \textit{MALAT1} for all lymphocytes or \textit{FTL} and \textit{FTH1} for all monocytes). Due to their expression in multiple cell clusters, {differential gene expression (DGE) does not identify these as highly ranked significant genes for one cell cluster}.
{
\textit{MALAT1} is a highly-expressed non-coding RNA, whose enrichment may be a feature of damaged or low-quality cells in some scRNA-seq data sets~\cite{alvarez2020enhancing, rindler2021single}. However, the biological differential expression of \textit{MALAT1} between immune cell types has also been previously reported, including increased expression in lymphocytes compared with monocytes~\cite{sookoian2018metastasis}. In any case, the increased expression of \textit{MALAT1} is certainly a feature shared by multiple lymphocyte clusters in the PBMC data set. 
Genes \textit{FTL} and \textit{FTH1} encode, respectively, the light and heavy chain of ferritin, which is a major protein involved in iron storage and homeostasis. It is highly expressed within myeloid (monocyte) cells, which can be further modulated by inflammatory stimuli~\cite{cohen2010serum, theurl2006dysregulated, zarjou2019ferritin}.}

\begin{figure}[h]
\includegraphics[width=\textwidth]{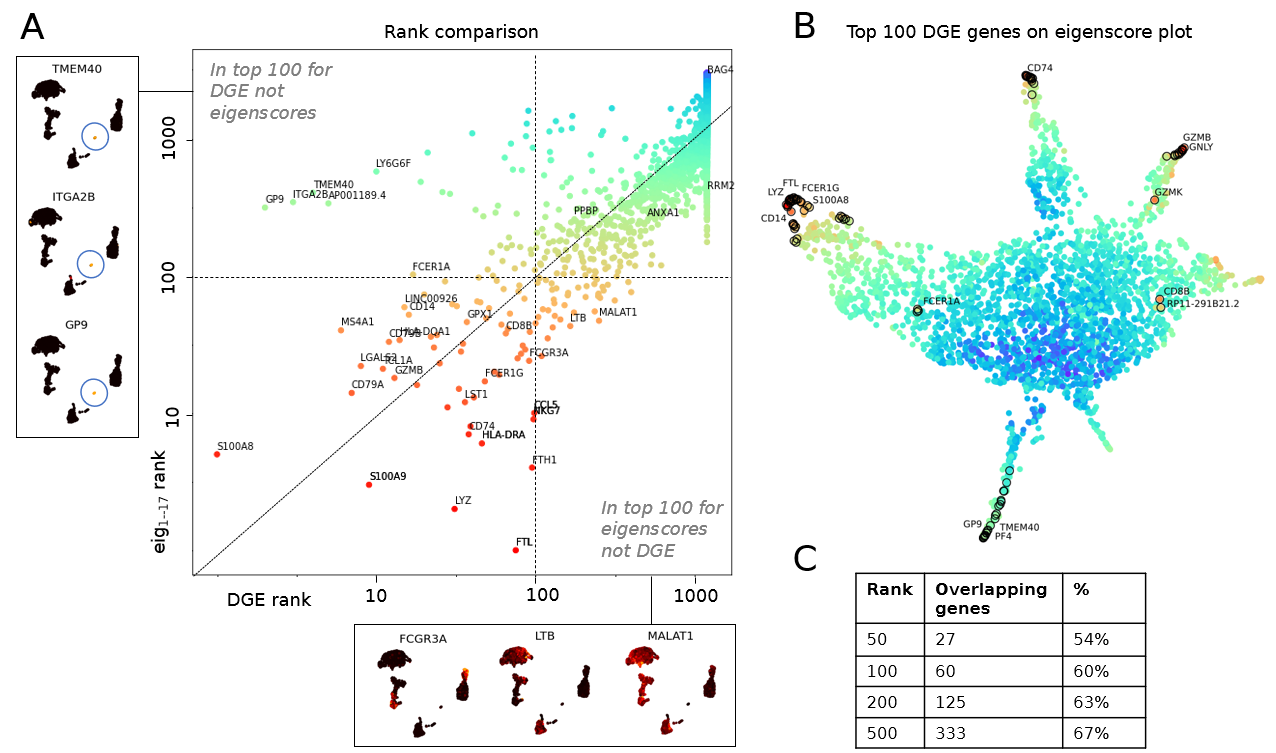}
	\caption{Eigenscores compared to differential gene expression (DGE) on PBMC data set~\cite{10XPBMC}. (\textbf{A}) Comparative study of DGE ranking using Seurat clustering {and a non-parametric Wilcoxon rank sum test} (log of rank computed from adjusted p-value on x-axis) versus ranking by norm in eigenscore space (log of eigenscore rank of 16 lowest frequencies on y-axis). Example genes in top 100 for one ranking but not the other shown on the sides.
		(\textbf{B}) Top 100 genes ranked by adjusted p-value in DGE marked on the eigenscore UMAP plot of genes from Figure \ref{fig:eigenvector_PBMC}. Two regions in the UMAP not found in the top of DGE are branch V from Figure \ref{fig:eigenvector_PBMC}B (T cell and lymphocyte genes that are expressed in larger groups of cells); branch VI (genes expressed in RRM2+ cluster that is not found by DGE).
		(\textbf{C}) Quantitative comparison of gene ranks given by adjusted p-value in DGE versus norm in 16-dimensional eigenscore space.
	}
	\label{fig:DE}
\end{figure}

{
While traditional DGE identifies markers unique to a given cluster or grouping of cells, eigenscores can also identify genes that exhibit biological variation within multiple disparate cell types. For example, the gene \textit{FCGR3A} can distinguish between major subpopulations of monocyte cells 
~\cite{pizzolato2019single, geng2021altered, cormican2020human}, major subpopulations of NK cells 
~\cite{victor2018epigenetic, crinier2021single} and major subpopulations of T cells
~\cite{pizzolato2019single}.
Eigenscore analysis ranks \textit{FCGR3A} highly (Figure \ref{Table:pbmc_eig}, $e_4$ and $e_5$), whereas with traditional DGE, it is not ranked in the top 10 genes for any cluster, which is perhaps due to its high level of expression across clusters in subtypes from disparate cell lineages.
}
Moreover, \textit{PPBP}, a highly differentially expressed marker gene on platelets, is ranked highly by eigenscores and not DGE. For completeness, we include a qualitative and quantitative comparison between eigenscore and DGE ranking {by non-parametric Wilcoxon rank sum test} (Figure \ref{fig:DE}) and present the set complement of the top scores. 

We can further explore the relationships between genes by projecting the gene space of eigenscores via UMAP (Figure~\ref{fig:eigenvector_PBMC}B). The visualisation of 16 dimensional low-frequency eigenscores (eigenscore 1--16 corresponding to $0<\lambda_i<0.1$) {emphasises} gene signals that are most coherent with the cell similarity graph structure.  We 
interpret {this visualisation of} gene space in Figure~\ref{fig:eigenvector_PBMC}B.
Genes plotting in the centre (blue) have low eigenscores,
where the gene expression is incoherent with the graph topology (see, for example, gene \textit{BAG4}). {Genes with similar expression patterns in the data set plot together, and a sequence of continuous transitions of gene signals plots continuously in eigenscore space (as illustrated in Figure \ref{fig:eig_explain}).}
The flares in the gene space with high eigenscores correspond to groups of genes strongly expressed on clusters corresponding to broad cell types or the cell cycle. We explore and interpret the continuous signal of gene space in Figure~\ref{fig:eigenvector_PBMC}B with flares interpreted clockwise as groups of genes expressed on: (I) platelets, (II) different subtypes of monocytes with continuous transition of genes, (III) B cells and dendritic cells, (IV) {cytotoxic lymphocytes (NK cells and CD8 T cells)}, (V) 
 {all lymphocytes but not myeloid cells (monocytes) (these are two separate developmental lineages)},
and (VI) {a previously unidentified subpopulation of cells within a larger cluster expressing a high level of cell cycle markers.}

{
Figure \ref{fig:DE}C gives a quantitative comparison of the amount of overlap in the eigenscore and DGE rankings, showing that there is consistent overlap while we also  find additional genes. We highlight that two flares of the gene space geometry are missed by DGE via non-parametric Wilcoxon rank sum test (see Figure \ref{fig:DE}B), particularly region VI. This region corresponds to genes involved with cell cycle progression, which are highest in a highly proliferative subpopulation of B cells in the PBMC data set. The genes in this region show strong overlap with gene signatures used by Seurat to identify the S phase (\textit{PCNA, TYMS, RRM2}) and G2/M phase (\textit{TOP2A, BIRC5, UBE2C, HMGB2, SMC4, CKS1B}). Notably, our eigenscore method is able to identify the cell cycle as an important source of biological variation in this data set that is missed by traditional cluster-based DGE without requiring the need to run an additional signature scoring method relying on the expression of a predefined set of markers for this biological phenomenon.}


\subsubsection{Eigenscores for Analysing Data with Continuous Structure: T Cells}

\begin{figure}[h]
 
    \includegraphics[width=\textwidth]{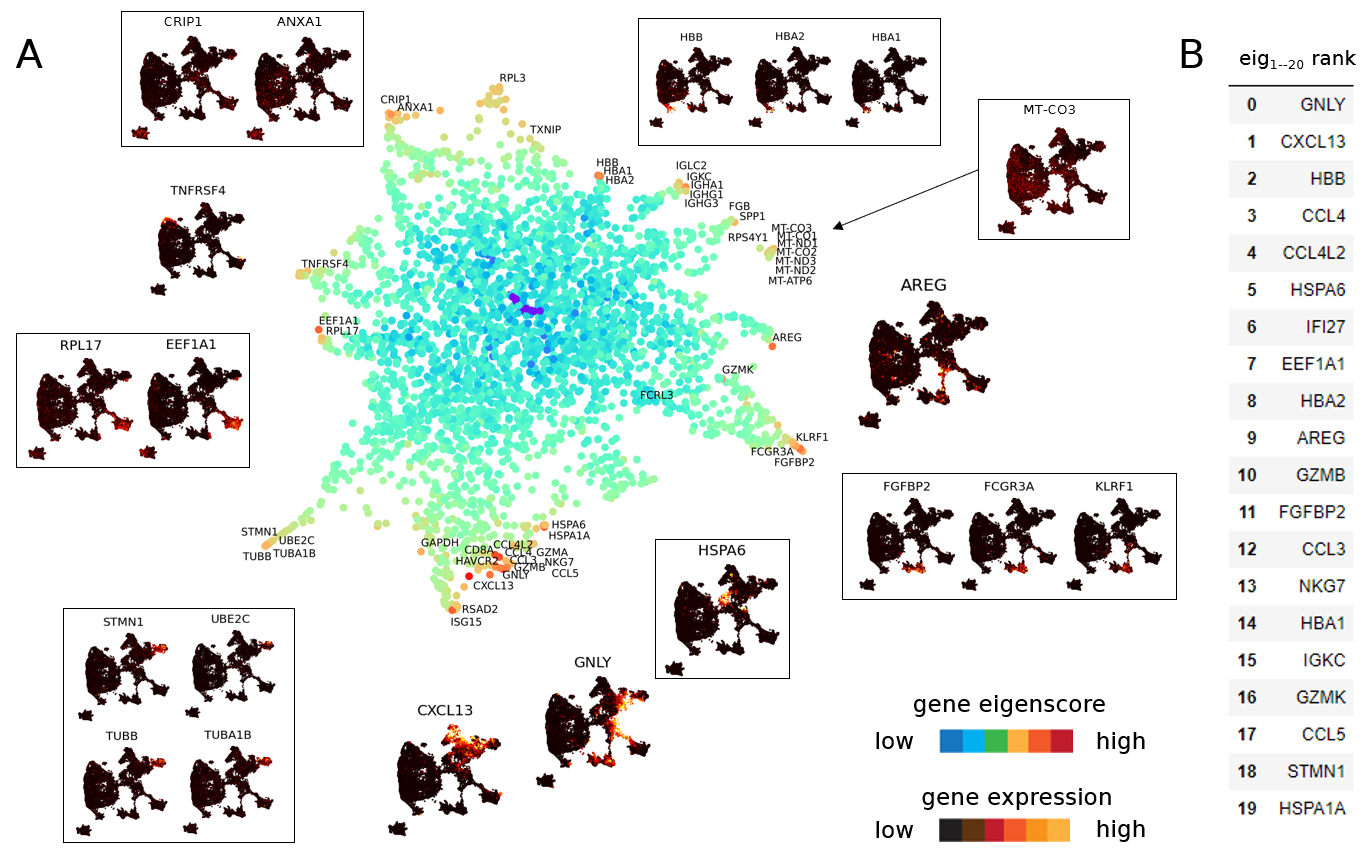}
    \caption{(\textbf{A}) UMAP of genes from T cell data set~\cite{lambrechts2018phenotype} in eigenscore space for eigenvectors 1--19, colour-coded for the logarithm of norm of the vector in 19-dimensional eigenscore space. Genes with similar expression group together and reveal substructure in the data set. Some genes have unique expression patterns not matched by other genes. 
    	Boxed genes represent a group of genes with similar expression whereas unboxed genes represent isolated gene behaviour.
    	(\textbf{B}) Top 20 genes ranked by norm in 1--19 dimensional eigenscore space.    
}
    \label{fig:eigenvector_TCell}
\end{figure}

We next compute eigenscores of the T cell data~\cite{lambrechts2018phenotype} (corresponding to $0<\lambda<0.1$).
As before, we can visualise the gene space in Figure \ref{fig:eigenvector_TCell}A where genes are near to another gene if they have similar expression patterns. For example, a large number of mitochondrial genes, such as \textit{MT-CO3}, that are lowly expressed {in cells across the data set} are grouped together as a distinct gene cluster.
{
The detection of cells expressing high levels of such mitochondrial genes can be an important step in filtering poor quality cells from scRNA-seq data sets~\cite{stegle2015computational}}. 

The continuous nature of the T cell data set is reflected in the many intermediate to high eigenscore genes that show coherent regions in the data set on multiple scales. While we can identify groups of cells that have coherent gene expression behaviour, such as the clusters formed by \textit{EEF1A1}+ cells, \textit{HBB}+ cells, \textit{ANXA1}+ cells and \textit{HSPA1A}+ cells, we also find genes that have unique expressions that are unlike any other gene signals (e.g., \textit{GNLY}{ and \textit{GZMB}, which are both known secreted cytotoxic effector genes found across T and NK cell subsets}). To explore the geometry of genes further, we analyse 
the top 20 genes ranked by eigenscore norm in 1--19 dimensional eigenscore space. We find \textit{AREG} (9th in eigenscore rank) does not show up on the DGE ranking. As shown in the UMAP cell subfigure, \textit{AREG} is expressed {\emph{in~between}} clusters of cell types, connecting NK cells and a cluster of CD8 T cells
(Figure \ref{fig:tcell_clusters}, see Section \ref{section:MLS_Tcell} for biological implications).
In this way, eigenscores provide insight into both the continuous and discrete nature of T cell behaviour.

\subsection{Multiscale Laplacian Score}\label{section:Results_MLS}

\begin{figure}[h!]
   
    \includegraphics[width=0.75\textwidth]{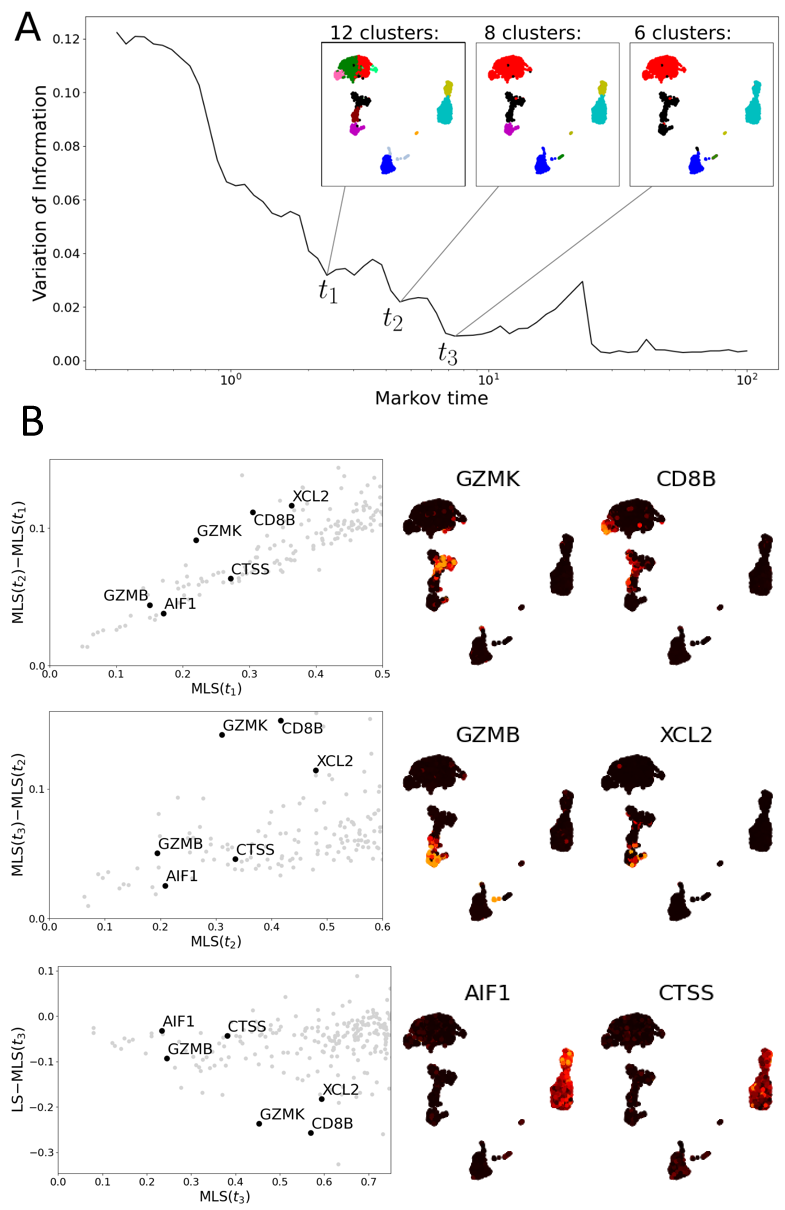}
    \caption{Multiscale Laplacian scores of PBMC data set~\cite{10XPBMC}. (\textbf{A}) The graph of variation of information of community structures returned by 100 iterations of the Louvain algorithm at each Markov time. Local minima indicate stable community structures and, hence, scales of interest. The community structures at three such minima are shown by colourings of UMAP plots. (\textbf{B}) Left: three scatter plots comparing the multiscale Laplacian scores of genes (grey dots) at successive times to one another (upper two) and of $t_3$ to the combinatorial Laplacian score (in all plots, axes are truncated). We highlight 6 genes of interest (annotated). Middle and Right: UMAP plots visualising the gene expression of six genes selected based on their MLS.}
    \label{fig:mls_pbmc}
\end{figure}

The multiscale Laplacian score (MLS), similar to the 0-dimensional combinatorial Laplacian score~\cite{govekClusteringindependentAnalysisGenomic2019} and gene connectivity score ~\cite{rizviSinglecellTopologicalRNAseq2017}, extends DGE to settings in which a stable partition of cells into groups is not feasible; therefore, no assignment of cells into groups is required. The MLS ranks genes by their consistency with the topological structure of the data set and performs such topological consistency analyses at multiple resolutions.

\begin{figure}[h!]
 
    \includegraphics[width=0.75\textwidth]{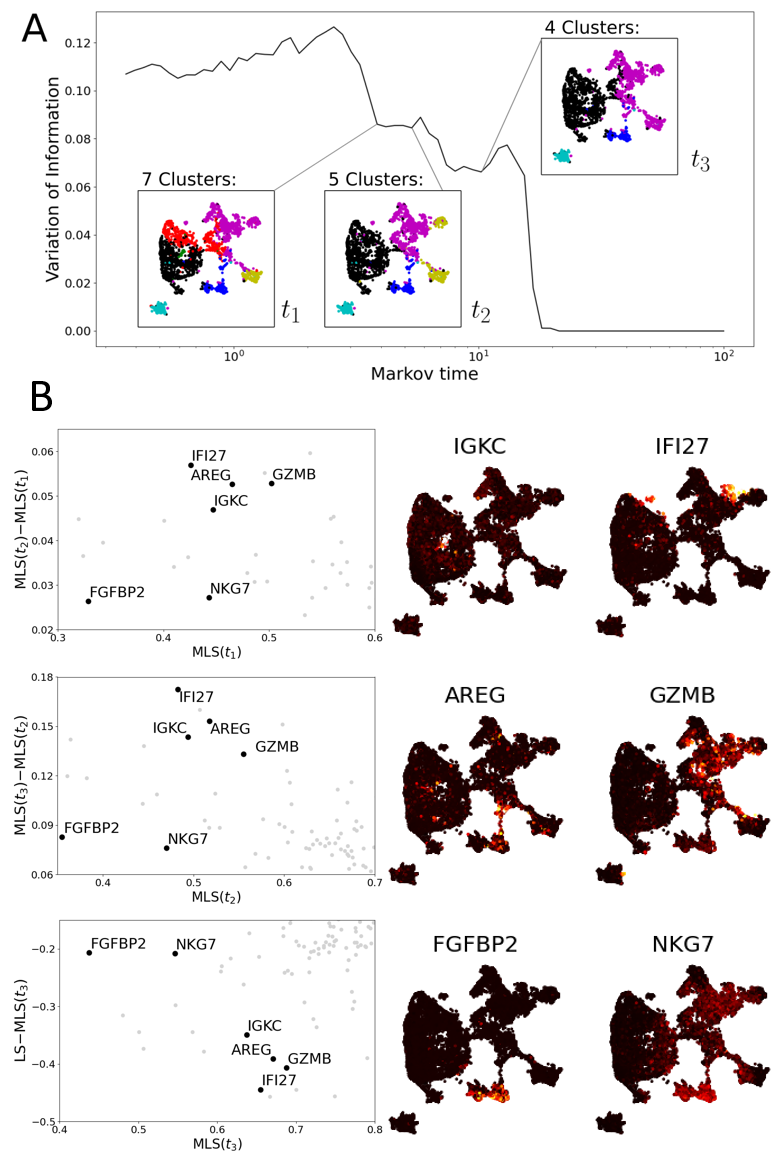}
    \caption{Multiscale Laplacian score of human T cell data set~\cite{lambrechts2018phenotype}. (\textbf{A}) The graph of variation of information of community structures.
    Again, local minima indicate scales of interest. Community structures at three scales are picked out. 
    (\textbf{B}) (\textbf{Left}): three scatter plots comparing the multiscale Laplacian scores of genes (grey dots) at successive times to one another (\textbf{left} and \textbf{middle} plot) and of $t_3$ to the combinatorial Laplacian score (in all plots, axes are truncated). We highlight 6 genes of interest (black dots; annotated). (\textbf{Middle} and \textbf{Right}): UMAP plots visualising the gene expression of six genes selected based on their MLS.}
    \label{fig:mls_tcell}
\end{figure}

The resolutions are determined by finding scales in the data that provide stable community structures. 
We reiterate that the MLS calculation does not use the obtained communities; rather, we use the resolution that provides a stable communities via local minima in variation of information (VI). 
{We highlight that the communities we find in both the PBMC and T cell data increase in size with increasing Markov time  (see \mbox{Figures \ref{fig:mls_pbmc} and \ref{fig:mls_tcell}}, panels A). This size increase illustrates
that the MLS at the selected Markov times tests for consistency at different scales in gene space.}

\subsubsection{Multiscale Laplacian Score of PBMC Data}
In Figure \ref{fig:mls_pbmc}, we apply MLS to the PBMC data set~\cite{10XPBMC}.
These data permit a stable clustering into five larger groups of cells. However, these clusters contain non-stable substructures (see Figure \ref{fig:mls_pbmc}A). The substructures largely align with the clusters found in the Seurat VST vignette~\cite{seuratVST} (Figure \ref{fig:pbmc_clusters}).
We find that the genes \textit{GZMK} and \textit{CD8B} exhibit a higher consistency with the structures at the first resolution ($t_1$) than at later resolutions (Figure \ref{fig:mls_pbmc}B). The gene \textit{GZMK} is highly expressed on the intersection of two communities at $t_1$ which correspond to naive and memory CD8 T cells, but it is not highly expressed on the union of these two communities (the two communities merge at resolution $t_2$). \textit{GZMK} seems to {mark} a transition between these two clusters. Similarly,
{
\textit{CD8B} expression is detected within the left-most community assigned to the broader CD4 T cell cluster of the UMAP plots} (Figure \ref{fig:pbmc_clusters}), which is merged into a larger community at $t_2$. The genes \textit{GZMB} and \textit{XCL2} are examples of features with low MLS at $t_2$ but relatively high MLS at $t_3$. The former is highly expressed on a community corresponding to effector CD8 T cells (cluster 6 in Figure \ref{fig:pbmc_clusters}), the latter on the intersection of the effector and the naive/memory T cell communities (clusters 6 and 5 and 7 in Figure \ref{fig:pbmc_clusters}). At resolution $t_3$, clusters 5 and 7 are merged with cluster 6. 
The communities at resolution $t_3$ correspond to the seven {populations} describing the different cell types with the NK and CD8 T cells merged.

Examples of genes with low MLS at $t_3$, relative to the standard LS as in~\cite{govekClusteringindependentAnalysisGenomic2019} and MLS at other resolutions, include \textit{AIF1} and \textit{CTSS}. Both genes are highly and consistently expressed on the community consisting of \textit{CD14}+ and \textit{FCGR3A}+ monocytes (Figure \ref{fig:pbmc_clusters})~\cite{seuratPBMC, seuratVST}. Resolution $t_3$ is the first resolution at which \textit{CD14}+ and \textit{FCGR3A}+ monocytes form a single community.

\subsubsection{Multiscale Laplacian Score of T Cell Data}\label{section:MLS_Tcell}
We compute the MLS to a human T cell data set from Labrechts et al.~\cite{lambrechts2018phenotype} (Figure \ref{fig:mls_tcell}).
As remarked by~\cite{govekClusteringindependentAnalysisGenomic2019}, these data do not allow for any partitioning into stable clusters ({see high VI values in  Figure \ref{fig:mls_tcell}A)}. Next, we determine three resolutions of interest based on the VI. Genes with a relatively low MLS at the finest resolution, $t_1$, include \textit{IGKC} and \textit{IFI27}. Both are highly expressed on a small group of cells (in the center of left hand side and top right of UMAP plot respectively; see Figure \ref{fig:mls_tcell}B). 

{
The gene \textit{IGKC} is an immunoglobulin gene, an antibody component found in B cell subsets, particularly plasma cells~\cite{lee2021single}.
Cells expressing \textit{IGKC} are also \textit{JCHAIN}+ and positive for antibody subtypes suggestive of class switching (e.g., \textit{IGHG1} and \textit{IGHA1}). Since this is a T cell data set, this almost certainly indicates that the cells in question are doublets (two cells in the same experimental droplet), specifically T cells binding B cells. While not representing single cell states, it is important that these readings are picked up in the analysis.}

{
The gene \textit{IFI27} is part of an antiviral/interferon-induced (IFI) response signature. 
It is particularly interesting that MLS can detect a specific transcriptional programme shared across multiple cell types (CD4+ and CD8+ T cells). 
This could represent a shared T cell programme directed against viruses or induced during stress responses (e.g., for scRNA-seq processing)~\cite{ullah2021antiviral}.}


{As a particular example of informative gene prioritisation, we highlight \textit{AREG} at resolution $t_2$, where it is expressed highly on a group of cells bridging nearby clusters identified as NK cells and CD8 T cells based on overall marker expression of each cluster (Figure \ref{fig:tcell_clusters}). Within the immune system, \textit{AREG} is expressed by subsets of NK cells and other types of innate lymphoid cells (ILCs)~\cite{crinier2021single, monticelli201533}, where it plays an important role in mediating type 2 immunity~\cite{monticelli201533, zaiss2015emerging}. Despite bridging different clusters in our global clustering and that of the original authors~\cite{lambrechts2018phenotype}, this population likely represents cells in various states of transition between two previously described \textit{AREG}+ NK cell phenotypes: one with high levels of secreted molecules associated with effector functions (\textit{CCL3, CCL4}) and the other expressing homing receptors associated with a more circulatory phenotype (\textit{CD44, SELL})~\cite{crinier2021single} (see Figure \ref{fig:extra_genes}).
Therefore, while the original authors identified these two subsets as discrete NK and type 1 ILC-like cell types, respectively (\cite{lambrechts2018phenotype} Figure S13), both our eigenscore and MLS-informed approaches highlight \textit{AREG} as a shared feature, supporting the notion that these two populations may be consistent with a more continuous transition between \textit{CCL3}+ and \textit{SELL}+ states within the NK cell population. This interpretation is further supported by the preserved expression of NK cell markers (e.g., CD94/\textit{KLRD1}, NKG2A/\textit{KLRC1}) in the \textit{SELL}+\textit{AREG}+ cells, which is often considered to be a feature of NK cells that is not shared by otherwise closely related type 1 ILCs~\cite{bennstein2020umbilical, bernink2013human}.}

Similarly, \textit{GZMB} is highly expressed on the intersection of exhausted and proliferating T cells (see Figure \ref{fig:tcell_clusters}), two clusters of which are visible in the community structure at $t_2$ but merge at $t_3$. Finally, at Markov time $t_3$, \textit{FGFBP2} and \textit{NKG7} are examples of genes with relatively low expression that are highly and consistently expressed on the cluster of {NK} cells.

\subsection{Persistent Rayleigh Quotient}\label{section:Results_PRQ}

Cell bifurcation methods, such as trajectory inference algorithms, seek to assign a pseudotime to each cell by fitting a tree onto the data set ~\cite{saelensComparisonSinglecellTrajectory2019, vandaeleStableTopologicalSignatures2021} or fit a statistical model to each gene and then test against a null model ~\cite{vandenbergeTrajectorybasedDifferentialExpression2020,
trapnellDynamicsRegulatorsCell2014,
qiuReversedGraphEmbedding2017,
lonnbergSinglecellRNAseqComputational2017,
jiTSCANPseudotimeReconstruction2016}.
The Rayleigh quotient and Laplacian score have proven useful in selecting genes~\cite{govekClusteringindependentAnalysisGenomic2019} but are agnostic to any prior cell knowledge or meta data. 
Here, we use additional time information to filter the graph.
{
This time information could be real developmental time 
or an inferred pseudotime derived from trajectory analysis.
Using the persistent Rayleigh quotient (PRQ)}, we can then separate 
genes that have different roles in this differentiation process.

\begin{figure}[h]
 
    \includegraphics[width=0.75\textwidth]{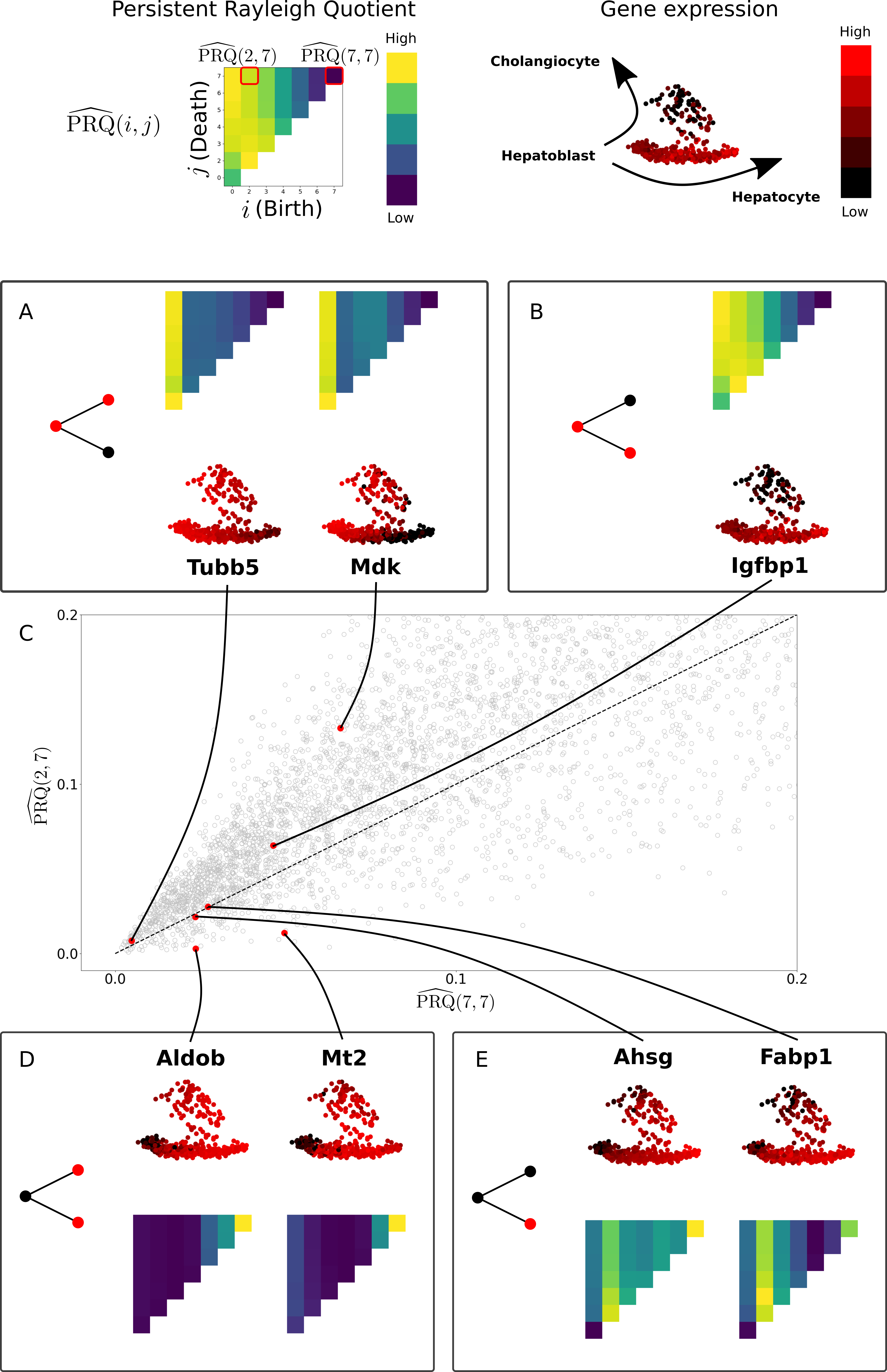}
    \caption{
The persistent Rayleigh quotient separates genes by their role in a cell differentiation process. The PRQ is parameterised by birth ($i$) and death ($j$), each pair $(i, j)$ assigning a non-negative number to every gene. We plot these values for each gene for $(i=7, j=7)$ on the x-axis and $(i=2, j=7)$ on the y-axis on subfigure (\textbf{C}). Selected for display (\textbf{A},\textbf{B},\textbf{D},\textbf{E}) are top differentially expressed genes from~\cite{muEmbryonicLiverDevelopmental2020}
on the data from~\cite{yangSinglecellTranscriptomicAnalysis2017} (see Figure~\ref{fig:mouse-data set}). Genes \textit{Tubb5}, \textit{Mdk}, and \textit{Igfbp1} are expressed in parent and one daughter cell lineage, hepatoblast to (\textbf{A}) cholangiocyte or (\textbf{B}) hepatocyte and lie above the diagonal. Genes \textit{Aldob} and \textit{Mt2} are expressed in both daughter cell types but not in the parent cell type (\textbf{D}), and they lie below the diagonal. Genes \textit{Ahsg} and \textit{Fabp1} are only expressed in one daughter cell type (\textbf{E}) and lie on the diagonal (compare with Figure \ref{fig:pl_explain}).
    }
\label{fig:pl_result_mouse_hepatic}
\end{figure}

We apply the PRQ to a bifurcation describing the differentiation of mouse hepatic cells by Yang et al.~\cite{yangSinglecellTranscriptomicAnalysis2017} (see Figure \ref{fig:pl_result_mouse_hepatic}).
Hepatoblasts, hepatocytes, and cholangiocytes were sampled from mouse embryos at seven time points, from embryonic day 10 to day 17.
Hepatoblasts are a parental cell type whose daughter cells differentiate into hepatocyte and cholangiocyte cell types.
As the topologically interesting direction is in `reverse time', we assign a filtration to the graph
by assigning a node from day \(t\) the filtration value \(17-t\).
{
For each pair of steps \(i\) and \(j\)
with \(i \leq j\)
in the filtration of the graph, we apply the 
normalised persistent Rayleigh quotient to a particular gene.
We represent this as a two-dimensional score for each gene in Figure \ref{fig:pl_result_mouse_hepatic}.
This is reminiscent of the birth--death persistence
diagram in TDA
where the x-axis records the filtration step
where certain topological features first appear, known as the birth time,
while the y-axis records where these features finally vanish, known as the death time.
In order to show the largest
differences between parts of the PRQ score for a gene,
we compare the normalised persistent Rayleigh quotients \(\NPRQ(2, 7)\) with \(\NPRQ(7,7)\)
in Figure \ref{fig:pl_result_mouse_hepatic}C.}

We have highlighted genes that were found to be differentially expressed during hepatoblast differentiation in~\cite{muEmbryonicLiverDevelopmental2020} (Figure \ref{fig:pl_result_mouse_hepatic}A,B,D,E).
As expected, genes such as \textit{Tubb5}, \textit{Mdk}, and \textit{Igfbp1},
which are expressed in hepatoblasts and only one of hepatocytes or cholangiocytes,
have a higher value for the persistent Rayleigh quotient than the full Rayleigh quotient
(Figure \ref{fig:pl_result_mouse_hepatic}A,B).
{
The gene \textit{Mdk} is known to show a decrease during hepatoblast maturation towards hepatocytes {in utero}, corresponding with the upregulation of genes involved in hepatocyte function including \textit{Aldob}~\cite{muEmbryonicLiverDevelopmental2020, su2017single}; however, this gene is preserved in cholangiocyte populations~\cite{mdkHPA}, agreeing with our observations using the PRQ.
\textit{Igfbp1} is not known to play a role in differentation into hepatocytes; however, the expression of \textit{Igfbp1} has been previously shown in hepatocytes, where it has a prosurvival role that can be enhanced by p53 activity~\cite{leu2007hepatic}.}
Genes \textit{Aldob} and \textit{Mt2} are expressed in cholangiocytes and hepatocytes but not hepatoblasts.
Hence,
their persistent Rayleigh quotient has a lower value than the full Rayleigh quotient (Figure \ref{fig:pl_result_mouse_hepatic}D).
Finally, the persistent Rayleigh quotient and full Rayleigh quotient for \textit{Fabp1}
and \textit{Ahsg} are almost the same. 
This corresponds to the fact that
\textit{Fabp1} and \textit{Ahsg} are highly expressed in only one of the daughter cell types
(Figure \ref{fig:pl_result_mouse_hepatic}E).
The full Rayleigh quotient \(\NPRQ(7, 7)\) can sort genes based on how coherently expressed
they are with respect to the underlying graph, but it does not distinguish between different expression patterns relevant to development. In contrast, the persistent Rayleigh quotient can differentiate genes whose expression pattern is relevant to bifurcation.

%% file: conclusions.tex
\section{Conclusions}

Inspired by the multiscale nature of topological data analysis, we proposed three multiscale methods relying on spectral graph theory and signal processing, which complement standard differential gene expression. We showcased the versatility of eigenscores and multiscale Laplacian scores (MLS) on different data sets.  These methods select genes in an unsupervised and continuous manner without requiring a clustering of cells {and therefore can identify genes with important biological variation within and across disparate clusters, which may be more difficult to identify using traditional DGE.}
The persistent Rayleigh quotient (PRQ) was applied to a cell differentiation data set, which validated a known cellular bifurcation and separated genes based on their role in the differentiation process. These methods proposed provide multiple different rankings of genes. Future directions include the systematic comparison of multiple rankings (e.g., using Hodge theory) {and summary statistics} to compare with methods that output one-dimensionally ranked genes. The PRQ also gives a rich representation for each gene, and future work will explore statistical integration with other pipelines. We provide available code, and a future goal is to create a topological genomics signaling package to increase accessibility and adoption.

While we focused on the geometry of gene space with a specific \(k\)-nn graph constructed using scRNA-seq, the proposed methods are flexible for other graphs, such as Mapper graphs~\cite{rizviSinglecellTopologicalRNAseq2017,rabadan2020identification}, but the resulting analysis would change if the underlying cell graph changes. The choice of resolution(s) for the MLS is not limited to Markov stability times (e.g.,  graph wavelets~\cite{tremblay2014graph}).
{Future directions include extending these signal selection approaches to other signals more generally (e.g., epigenetic factors),} other complex single-cell network structures~\cite{jeitziner2017two} or other higher-order networks~\cite{schaub2021signal,bick2021higher}, with a view towards data integration~\cite{kuchroo2021multimodal}. 

{The methods we present in this paper provide an exciting foundation to rethink the \textit{de novo} identification of important genes in scRNA-seq data sets. For example, by assessing the geometry of gene space using eigenscores, we are able to implement a new variant of gene set enrichment analysis, where the identification of meaningful groups of genes is driven by the expression patterns of the genes themselves at any scale, instead of a test statistic averaged across a predefined comparison of clusters. This approach allowed us to not only identify broad markers of lineages and cell types in our example data sets but also important smaller-scale biological phenomena such as the cell cycle and preserved \textit{AREG} expression bridging NK cell subtypes without relying on the need for optimal cluster assignment used by traditional DGE or biological priors required by signature scoring methods. In this way, our methods allow the user to build understanding of their data sets at multiple levels in a single analysis by identifying the key genes that drive these multiscale sources of biological variation.}

%% file: additional_figures.tex
\begin{appendix}
\section*{Appendix}
\begin{figure}[h]
	
	\includegraphics[width=0.65\textwidth]{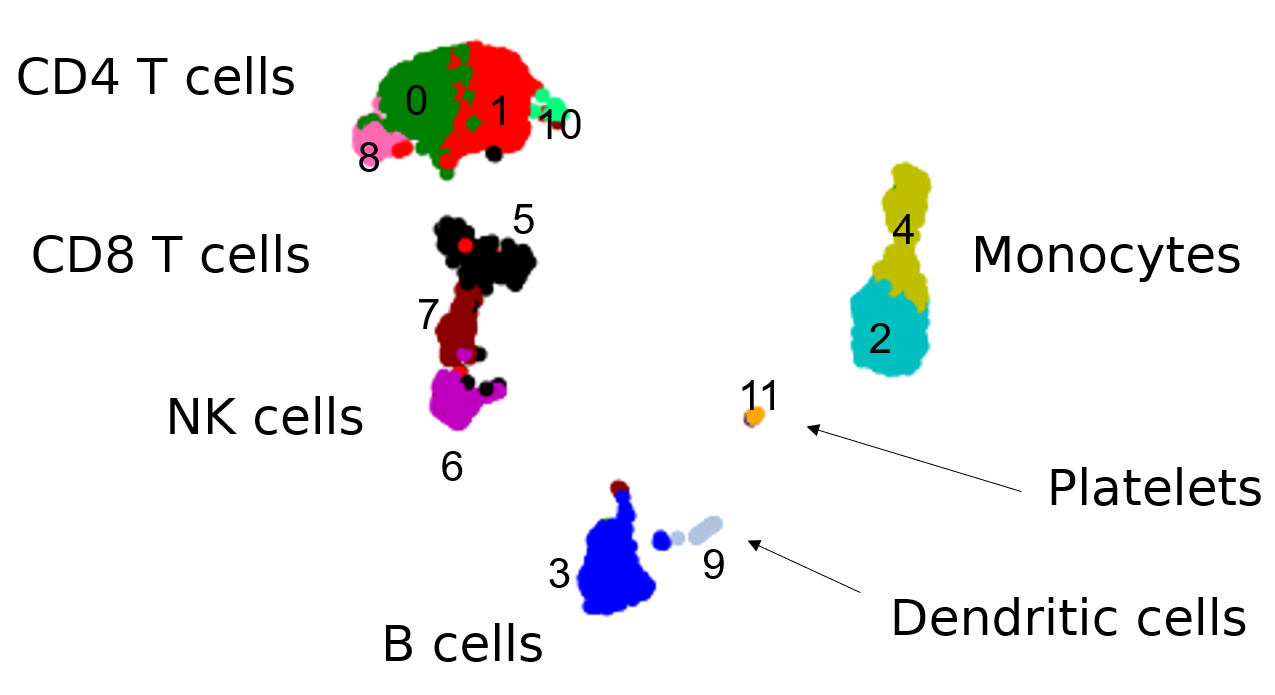}
	\caption{Seurat clusters on PBMC {data}~\cite{10XPBMC} from Seurat VST Vignette~\cite{seuratVST} numbered according to the vignette and interpretations of overarching cell types inferred from previous results. The cells in this data set divide into broad clusters corresponding to the cell types found in peripheral blood mononuclear cells: lymphocytes (T cells, NK cells, B cells), monocytes, and dendritic cells, as also platelets which are not mononuclear but are found in this specific data set. The DGE analysis from Seurat {(non-parametric Wilcoxon rank sum test~\cite{seuratDGE})} defines twelve smaller clusters, in particular sublcustering T cells, NK cells and monocytes, and searches only for differentially expressed genes on these subclusters.  
	}  
	\label{fig:pbmc_clusters}
\end{figure}

\begin{figure}[h]
	\includegraphics[width=\textwidth]{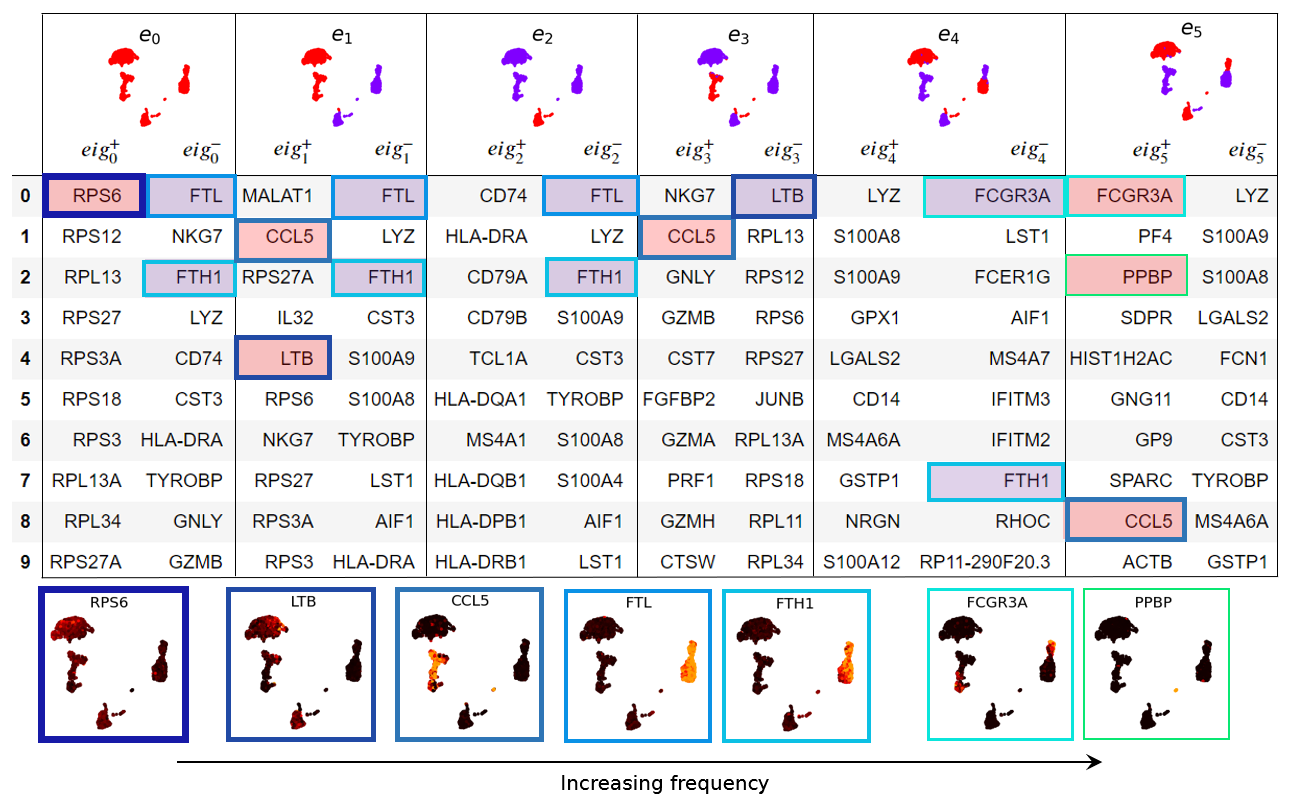}
	\caption{Eigenscore ranks for PBMC data~\cite{10XPBMC}. On the top row are plots of the Laplacian eigenvectors, coloured by sign (red positive, purple negative). For each eigenvector $e_i$, genes are listed with the highest alignment ($eig_i^+$) and highest anti-alignment ($eig_i^-$) with $e_i$. Below the table are a selection of genes ranked highly by eigenscores. For example gene \textit{FTL} shown below the table is strongly expressed on the monocyte cluster on the right, which is purple (negative) for both $e_1$ and $e_2$, hence \textit{FTL} has high scores on $eig_1^-$ and $eig_2^-$.
	}
		\label{Table:pbmc_eig}
\end{figure}

\begin{table}[h]
      \caption{Differentially expressed genes for PBMC data found by Seurat.}
    \label{Table:pbmc_dge}
    \adjustbox{scale=0.6,center}{
	\begin{tabular}{lllllllllllll}
		& 0     & 1      & 2      & 3         & 4             & 5    & 6      & 7      & 8             & 9        & 10      & 11         \\
		0 & RPS27 & LTB    & S100A8 & CD79A     & IFITM3        & GZMK & GZMB   & GZMH   & CD8B          & FCER1A   & IFIT1   & GP9        \\
		1 & RPL32 & IL32   & LGALS2 & MS4A1     & RP11-290F20.3 & CCL5 & FGFBP2 & CST7   & RP11-291B21.2 & ENHO     & IFIT3   & ITGA2B     \\
		2 & RPS6  & IL7R   & S100A9 & TCL1A     & LST1          & NKG7 & SPON2  & NKG7   & CD8A          & CLEC10A  & RTP4    & TMEM40     \\
		3 & RPS12 & CD3D   & CD14   & CD79B     & AIF1          & LYAR & GNLY   & CCL5   & S100B         & SERPINF1 & SPATS2L & AP001189.4 \\
		4 & RPL31 & AQP3   & FCN1   & HLA-DQA1  & MS4A7         & GZMA & PRF1   & GZMA   & CARS          & CD1C     & DDX58   & LY6G6F     \\
		5 & RPS14 & LDHB   & TYROBP & LINC00926 & IFI30         & IL32 & XCL2   & FGFBP2 & RPS12         & CACNA2D3 & RSAD2   & sep-05     \\
		6 & RPS25 & CD2    & MS4A6A & VPREB3    & CD68          & CD8A & AKR1C3 & CD8A   & RPL13         & HLA-DQB2 & MX1     & HGD        \\
		7 & LDHB  & CD40LG & LYZ    & HLA-DQB1  & FCER1G        & CTSW & CLIC3  & GZMB   & RPS6          & HLA-DQA2 & ISG15   & PTCRA      \\
		8 & RPS3A & TPT1   & GPX1   & CD74      & CFD           & CST7 & KLRD1  & CTSW   & CCR7          & HLA-DQA1 & IFI6    & TREML1     \\
		9 & RPL30 & CD3E   & CST3   & HLA-DRA   & SERPINA1      & HOPX & CST7   & CCL4   & RPL32         & NDRG2    & HERC5   & ITGB3     
\end{tabular}}
\end{table}

\begin{figure}[h]
	\includegraphics[width=\textwidth]{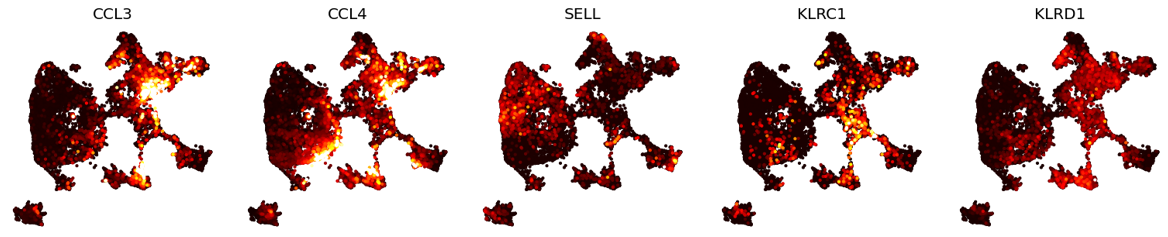}
	\caption{{
			Expression of relevant marker genes in the T cell data set.}}
	\label{fig:extra_genes}
\end{figure}

\vspace{-6pt}

\begin{figure}[h]
 
    \includegraphics[width=0.45\textwidth]{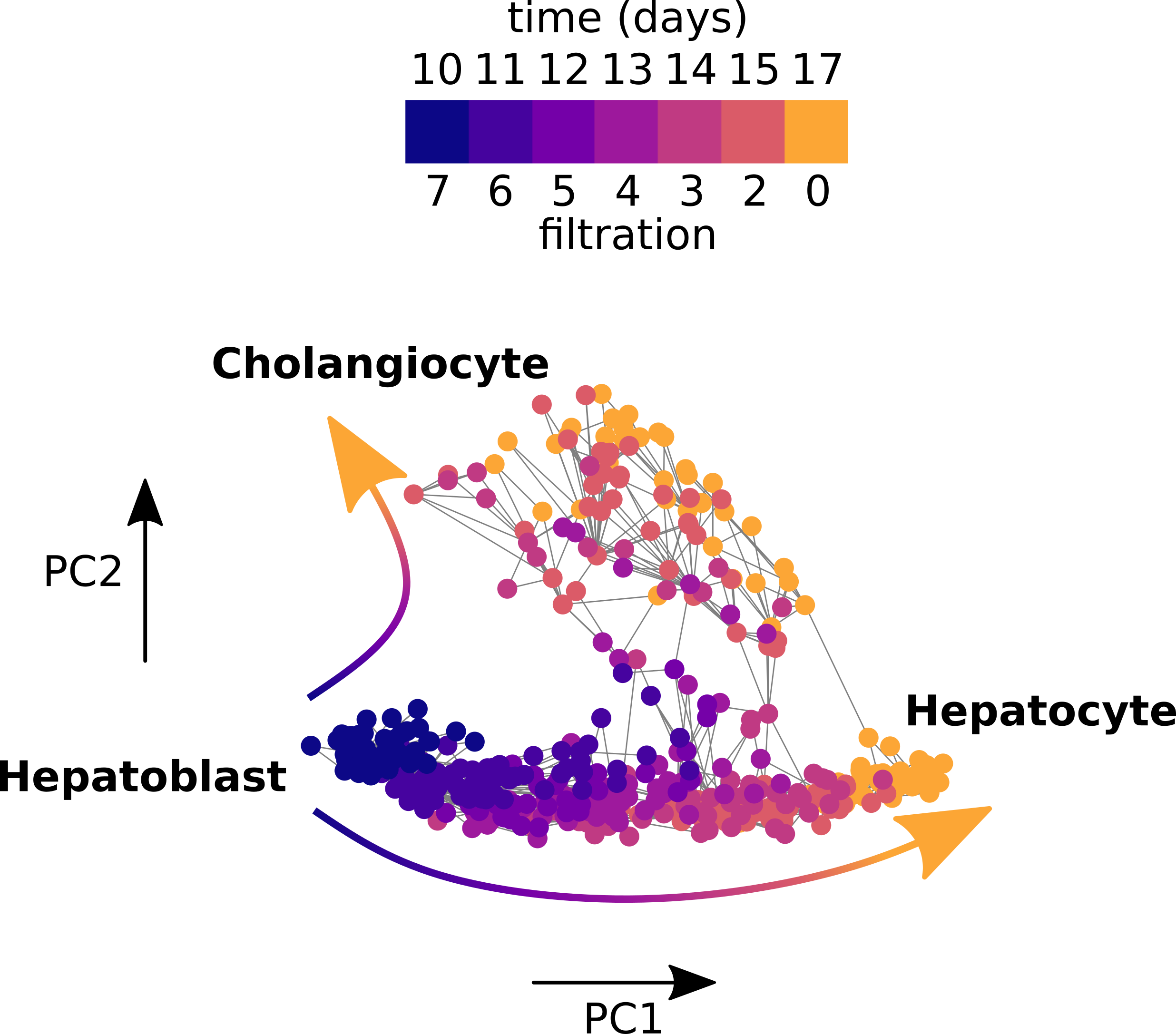}
    \caption{A weighted graph constructed from mouse foetal liver cells sampled from days 10--17 during development. Parent cell type hepatoblasts differentiate into two daughter cell types, cholangiocytes and hepatocytes.}
    \label{fig:mouse-data set}
\end{figure}
\end{appendix}